\documentclass{aa}
\usepackage{natbib}
\usepackage{graphics}
\usepackage{amssymb}
\newcommand{\fno}[1]{$^{(#1)}$}
\begin{document}
\title{Physical structure and CO abundance of low-mass protostellar
envelopes}
\author{J.K. J{\o}rgensen \and F.L. Sch\"{o}ier \and E.F. van
Dishoeck}
\institute{Leiden Observatory, P.O. Box 9513, NL-2300 RA Leiden, The
Netherlands}
\offprints{Jes K.\,J{\o}rgensen}
\mail{joergensen@strw.leidenuniv.nl} 
\date{Received: 14 December 2001 / Accepted: 2 May 2002}

\abstract{We present 1D radiative transfer modelling of the envelopes
of a sample of 18 low-mass protostars and pre-stellar cores with the
aim of setting up realistic physical models, for use in a chemical
description of the sources. The density and temperature profiles of
the envelopes are constrained from their radial profiles obtained from
SCUBA maps at 450 and 850~$\mu$m and from measurements of the source
fluxes ranging from 60 $\mu$m to 1.3 mm. The densities of the
envelopes within $\sim$10000 AU can be described by single power-laws
$\rho\propto r^{-\alpha}$ for the class 0 and I sources with $\alpha$
ranging from 1.3 to 1.9, with typical uncertainties of $\pm$0.2. Four
sources have flatter profiles, either due to asymmetries or to the
presence of an outer constant density region. No significant
difference is found between class 0 and I sources. The power-law fits
fail for the pre-stellar cores, supporting recent results that such
cores do not have a central source of heating. The derived physical
models are used as input for Monte Carlo modelling of submillimeter
C$^{18}$O and C$^{17}$O emission. It is found that class I objects
typically show CO abundances close to those found in local molecular
clouds, but that class 0 sources and pre-stellar cores show lower
abundances by almost an order of magnitude implying that significant
depletion occurs for the early phases of star formation. While the
2--1 and 3--2 isotopic lines can be fitted using a constant fractional
CO abundance throughout the envelope, the 1--0 lines are significantly
underestimated, possibly due to contribution of ambient molecular
cloud material to the observed emission. The difference between the
class 0 and I objects may be related to the properties of the CO ices.

}
\maketitle

\section{Introduction}
In the earliest, deeply-embedded stage a low-mass protostar is
surrounded by a collapsing envelope and a circumstellar disk through
which material is accreted onto the central star, while the envelope
is dissipated simultaneously through the action of the powerful jets
and outflows driven by the young star. Traditionally, young stellar
objects (YSOs) have been classified according to their spectral energy
distributions (SEDs) in the class I-III scheme \citep{lada87,adams87}
describing the evolution of YSOs from the young class I sources to the
more evolved pre-main sequence class III sources. This classification
scheme was further expanded by \cite{andre93} to include sources that
mainly radiate at submillimeter wavelengths (i.e., with high ratios of
their submillimeter and bolometric luminosities, $L_{\rm sub mm}/L_{\rm
bol}$) and it was suggested that these so-called class 0
sources correspond to the youngest deeply embedded protostars. Even
earlier in this picture of low-mass star formation, the starless cores
of \cite{myers83} and \cite{benson89} are good candidates for
pre-stellar cores, i.e., dense gas cores that are on the brink of
collapse and so leading to the class 0 and I phases.

The \cite{shu77} model predicts that the outer parts of the envelope
follow a $\rho\propto r^{-2}$ density profile similar to the solution
of an isothermal sphere \citep{larson69}, while within the collapse
radius, which is determined by the sound speed in the envelope
material and the time since the onset of the collapse, the density
tends to flatten nearing a $\rho\propto r^{-1.5}$ profile in the
innermost parts. This model has subsequently been refined to include
for example rotational flattening \citep{terebey84} where such effects
are described as a perturbation to the \cite{shu77} solution.

An open question about the properties of YSOs in the earliest
protostellar stage is how the structure of the envelope reflects the
initial protostellar collapse and how it will affect the subsequent
evolution of the protostar, for example in defining the properties of
the circumstellar disk from which planets may be formed later. One of
the possible shortcomings of the Shu-model is the adopted, constant,
accretion rate. \cite{foster93} performed hydrodynamical simulations
of the stages before the protostellar collapse and found that the
structure of the core at the collapse initiating point is highly
dependent on the initial conditions; only in the case where a large
ratio exists between the radius of an outer envelope with a flat
density profile and an inner core with a steep density profile, will
the core evolve to reproduce the conditions in the Shu model. In an
analytical study, \cite{henriksen97} suggested that the accretion
history of protostars could be divided into two phases for cores with
a flat inner density profile: a violent early phase with high
accretion rates (corresponding to the class 0 phase) that declines
until a phase with mass accretion rates similar to the predictions in
the Shu-model is reached (class I objects), i.e., a distinction
between class 0 and I objects based on ages. Whether this is indeed
the case has recently been questioned by \cite{jayawardhana01}, who
instead suggest that both class 0 and I objects are protostellar in
nature, but just associated with environments of different physical
properties, with the class 0 objects in more dense environments
leading to the higher accretion rates observed towards these sources.

The chemical composition of the envelope may be an alternative tracer
of the evolution. Indeed, for high-mass YSOs, combined infrared and
submillimeter data have shown systematic heating trends reflected in
the ice spectra, gas/ice ratios and gas-phase abundances
\citep[e.g.,][]{gerakines99,boogert00,vandertak00vdc}. One of the
prime motivations for this work is to extend similar chemical studies
to low-mass objects, and extensive (sub-)millimeter line data for a
sample of such sources are being collected at various telescopes,
which can be complemented by future SIRTF infrared data.

In order to address these issues the physical parameters within the
envelope, in particular the density and temperature profiles and the
velocity field, are needed. The first two can be obtained through
modelling of the dust continuum emission observed towards the sources,
while observations of molecules like CO and CS can trace the gas
component and velocity field. At the densities observed in the inner
parts of the envelopes of YSOs it is reasonable to expect gas and dust
coupling, which is usually expressed by the canonical dust-to-gas
ratio of 1:100 and the assumption that the dust and gas temperatures
are similar. Therefore a physical model for the envelopes derived on
the basis of the dust emission can be used as input for modelling of
the abundances of the various molecules.

Recently, \cite{shirley00} and \cite{motte01} have undertaken surveys
of the continuum emission of low-mass protostars - using respectively
SCUBA (at 450 and 850~$\mu$m) and the IRAM bolometer at 1.3 mm. Both
groups analyzed the radial intensity profiles (or brightness profiles)
for the individual sources, assuming that the envelopes are optically
thin, in which case the temperature follows a power-law dependence
with radius in the Rayleigh-Jeans limit. Assuming that the underlying
density distribution is also a power-law (i.e., of the type $\rho
\propto r^{-\alpha}$), one can then derive a relationship between the
radial intensity observed in continuum images and the envelope radius,
which will also be a power-law with an exponent depending on the
power-law indices of the density and temperature distributions. Both
groups find that the data sets are consistent with $\alpha$ in the
range 1.5--2.5 in agreement with previous results and the model
predictions. However, as both groups also notice, in the case where
the assumption about an optically thin envelope breaks down, the
temperature distribution and so the derived density distribution may
not be correctly described in this approach.

To further explore these properties of the protostellar envelopes we
have undertaken full 1D radiative transfer modelling of a sample of
protostars and pre-stellar cores (see Sect.~\ref{oursample}) using the
radiative transfer code DUSTY \citep{ivezic97}. Assuming power-law
density distributions we solve for the temperature distribution and
constrain the physical parameters of the envelopes by comparison of
the results from the modelling to SCUBA images of the individual
sources and their spectral energy distributions (SEDs) using a
rigorous $\chi^2$ method. Besides giving a description of the physical
properties of low-mass protostellar envelopes, the derived density and
temperature profiles are essential as input for detailed chemical
modelling of molecules observed towards these objects. Also, a good
description of the envelope structure is needed to constrain the
properties of the disks in the embedded phase
\citep[e.g.,][]{keene90,hogerheijde98, hogerheijde99, looney00}.

In Sect.~\ref{ourdata} our sample of sources is presented and the
reduction and calibration briefly discussed \citep[see
also][]{schoier02}. In Sect.~\ref{modelling} the modelling of the
sources is described and the derived envelope parameters
presented. The properties of the individual sources are described in
Sect.~\ref{indsources}. In Sect.~\ref{discuss} the implication of
these results are discussed and compared to other work done in this
field. The results of the continuum modelling will be used in a later
paper as physical input for detailed radiative transfer modelling of
molecular line emission for the class 0 objects - as has been done for
class I objects \citep{hogerheijde98} and high-mass YSOs
\citep{vandertak00} and as presented for the low-mass class 0 object
IRAS 16293-2422 \citep{ceccarelli00a,ceccarelli00b,schoier02}. In
Sect.~\ref{gasmodelling}, the first results of this radiative transfer
analysis for C$^{18}$O and C$^{17}$O is presented.

\section{Data, reduction and calibration}\label{ourdata}
\subsection{The sample}\label{oursample}
The class 0 sources in the sample have been chosen from the list of
\cite{andreppiv}, with the requirement that they should have a
distance of less than 450~pc, luminosity less than 50 L$_\odot$ and be
visible from the JCMT. In addition, \object{CB244} from the
\cite{shirley00} sample was included. These objects were supplemented
with two pre-stellar cores \object{L1689B} and \object{L1544}, also
from \cite{shirley00} and two class I objects, \object{L1489} and TMR1
taken from \cite{hogerheijde97,hogerheijde98}. To enlarge the class I
sample, \object{L1551-I5}, \object{TMC1A} and \object{TMC1} were
included as well. The physical properties for these three sources were
modelled using the same approach as the remainder of the sources,
based on SCUBA archive data. They were, however, not included in the
JCMT line survey, so the line modelling (Sect.~\ref{gasmodelling}) was
mainly based on data presented in the literature, in particular
\cite{hogerheijde98} and \cite{ladd98}.

For the class 0 objects we have adopted luminosities and distances
from \cite{andreppiv}, for the class I objects the values from
\cite{motte01} and for the pre-stellar cores and \object{CB244}
distances and luminosities from \cite{shirley00}. There are a few
exceptions, however: for the objects related to the Perseus region we
assume a distance of 220~pc and scale the luminosities from
\cite{andreppiv} accordingly, while a distance of 325~pc is assumed
for \object{L1157} as in \cite{shirley00}. The sample is summarized in
Table~\ref{sourcetab}. The class 0 object IRAS 16293-2422 treated in
\cite{schoier02} has been included for comparison here as well.
\begin{table*}
\caption{Sample of sources.}\label{sourcetab}
\begin{center}
\begin{tabular}{llllllll} \hline\hline
 & $\alpha(2000)$ & $\delta(2000)$ & $T_{\rm bol}$ & $L_{\rm bol}$ & $d$ & Other names & Type \\ 
                 & (hh mm ss) & (dd mm ss) & (K) & ($L_\odot$) & (pc) &  \\ \hline
\object{L1448-I2}        & 03 25 22.4 &  \parbox{0.3cm}{+}30 45 12 & 60  & 3   & 220 & 03222+3034       & Class 0     \\
\object{L1448-C}         & 03 25 38.8 &  \parbox{0.3cm}{+}30 44 05 & 54  & 5   & 220 & L1448-MM         &             \\
\object{N1333-I2}        & 03 28 55.4 &  \parbox{0.3cm}{+}31 14 35 & 50  & 16  & 220 & 03258+3104/SVS19 &             \\
\object{N1333-I4A}       & 03 29 10.3 &  \parbox{0.3cm}{+}31 13 31 & 34  & 6   & 220 &                  &             \\
\object{N1333-I4B}       & 03 29 12.0 &  \parbox{0.3cm}{+}31 13 09 & 36  & 6   & 220 &                  &             \\
\object{L1527}           & 04 39 53.9 &  \parbox{0.3cm}{+}26 03 10 & 36  & 2   & 140 & 04368+2557       &             \\
\object{VLA1623}         & 16 26 26.4 & \parbox{0.3cm}{--}24 24 30 & 35  & 1   & 160 &                  &             \\
\object{L483}            & 18 17 29.8 & \parbox{0.3cm}{--}04 39 38 & 52  & 9   & 200 & 18148-0440       &             \\
\object{L723}            & 19 17 53.7 &  \parbox{0.3cm}{+}19 12 20 & 47  & 3   & 300 & 19156+1906       &             \\
\object{L1157}           & 20 39 06.2 &  \parbox{0.3cm}{+}68 02 22 & 42  & 6   & 325 & 20386+6751       &             \\
\object{CB244}           & 23 25 46.7 &  \parbox{0.3cm}{+}74 17 37 & 56  & 1   & 180 & 23238+7401/L1262 &             \\
\object{IRAS 16293-2422} & 16 32 22.7 & \parbox{0.3cm}{--}24 28 32 & 43  & 27  & 160 &                  & \fno{a}    \\\hline
\object{L1489}           & 04 04 43.0 &  \parbox{0.3cm}{+}26 18 57 & 238 & 3.7 & 140 & 04016+2610       & Class I     \\
\object{TMR1}            & 04 39 13.7 &  \parbox{0.3cm}{+}25 53 21 & 144 & 3.7 & 140 & 04361+2547       &             \\
\object{L1551-I5}        & 04 31 34.1 &  \parbox{0.3cm}{+}18 08 05 &  97 & 28  & 140 & 04287+1801       & \fno{b}    \\
\object{TMC1A}           & 04 39 34.9 &  \parbox{0.3cm}{+}25 41 45 & 172 & 2.2 & 140 & 04365+2535       & \fno{b}    \\
\object{TMC1}            & 04 41 12.4 &  \parbox{0.3cm}{+}25 46 36 & 139 & 0.66& 140 & 04381+2540       & \fno{b}    \\\hline
\object{L1544}           & 05 04 17.2 &  \parbox{0.3cm}{+}25 10 44 & 18  & 1   & 140 &                  & Pre-stellar \\
\object{L1689B}          & 16 34 49.1 & \parbox{0.3cm}{--}24 37 55 & 18  & 0.2 & 160 &                  &             \\\hline
\end{tabular}
\end{center}
Notes: \fno{a}Class 0 object treated in
\cite{schoier02}. \fno{b}Class I object not included in the JCMT line
survey but with CO observations from \cite{hogerheijde98,ladd98}.
\end{table*}

\subsection{Submillimeter continuum data}
Archive data obtained from the Submillimetre Common-User Bolometer
Array, (SCUBA), on the James Clerk Maxwell Telescope\footnote{The JCMT
is operated by the Joint Astronomy Centre in Hilo, Hawaii on behalf of
the parent organisations: the Particle Physics and Astronomy Research
Council in the United Kingdom, the National Research Council of Canada
and the Netherlands Organization for Scientific Research.} (JCMT), on
Mauna Kea, Hawaii were adopted as the basis for the analysis. Using
the 64 bolometer array in jiggle mode, it is possible to map a
hexagonal region with a size of approximately 2.3\arcmin\
simultaneously at, e.g., 450~$\mu$m and 850~$\mu$m. It is also
possible to combine jiggle maps with various offsets to cover a larger
region. To perform the initial reduction of the data, the package SURF
\citep{surfman} was used following the description in
\cite{scubacookbook}. The individual maps were extinction corrected
with measurements of the sky opacity $\tau$ obtained at the Caltech
Submillimeter Observatory (CSO) and using the relations from
\cite{taucalib} to convert the CSO 225 GHz opacity to estimates for
the sky opacity at 450~$\mu$m and 850~$\mu$m. The sky opacities can
also be estimated using skydips, and in cases where these were
obtained, the two methods agreed well. Most of the sources were
observed in the course of more than one program and on multiple days,
so wherever possible available data obtained close in time were used,
coadding the images to maximize the signal-to-noise and field
covered. In the coadding, it is possible to correct for variations in
the pointing by introducing a shift for each image found by, e.g.,
fitting gaussians to the central source. We chose, however, not to do
this, because only minor corrections were found between the individual
maps. Two sources, \object{L1157} and \object{CB244} only had usable
data at 850~$\mu$m \cite[see also][]{shirley00}, so supplementary data
for these two sources were obtained in October 2001 at 450~$\mu$m (see
Sect.~\ref{newcalib}).

For each source the flux scale was calibrated using available data for
one of the standard calibrators, either a planet or a strong
submillimeter source like CRL618. From the calibrated maps the total
integrated fluxes were derived and the 1D brightness profiles were
extracted by measuring the flux in annuli around the peak flux. The
annuli were chosen with radii of half the beam (4.5\arcsec\ for the
450~$\mu$m data and 7.5\arcsec\ for the 850~$\mu$m data) so that a
reasonable noise-level is obtained, while still making the annuli
narrow enough to get information about the source structure without
oversampling the data. Actually the spread in the fluxes measured for
the points in each annulus due to instrumental and calibration noise
was negligible compared to the spread due to (1) the gradient in
brightness across each annuli, and (2) deviations from circular
structure of the sources. One problem in extracting the brightness
profiles was presented by cases where nearby companions were
contributing significantly when complete circular annuli were
constructed - the most extreme example being N1333-I4 with two close
protostars. In these cases emission from ``secondary'' components was
blocked out by simply not including data-points in the direction of
these closeby sources when calculating the mean flux in each annulus.

\subsection{SCUBA observations of L1157 and CB244}\label{newcalib}
The observations of \object{L1157} and \object{CB244} were obtained on
October 9th, 2001. Calibrations were performed by observing Mars and
the secondary calibrator, CRL2688, immediately before the
observations. Skydips were obtained immediately before the series of
observations (all obtained within 3 hours) giving values for the sky
opacity of $\tau_{450}=1.2$ and $\tau_{850}=0.23$, which agree well
with the sky opacities estimated at the CSO during that night. From
gaussian fits to the central source the conversion factor from the V
onto the Jy beam$^{-1}$ scale ($C_\lambda$) was estimated and is
summarized in Table~\ref{caliboverview} together with the beam size
$\theta_{\rm mb}$ also estimated from the gaussian fit to the
calibration source. For Mars the estimate of the beam size was
obtained by deconvolution with the finite extent of the planet, while
CRL2688 was assumed to be a point source \citep{sandell94}.
\begin{table}
\caption{Summary of the calibration for the October 2001
data.}\label{caliboverview}
\begin{center}
\begin{tabular}{llllll}\hline\hline
        &$\lambda$   & $\tau$\fno{a} & $C_\lambda$\fno{b}& $\theta_{\rm mb}$\fno{c} \\\hline
Mars    & 450~$\mu$m & 1.2    & 395.1       & 8.9\arcsec       \\
CRL2688 & 450~$\mu$m & 1.2    & 310.5       & 9.1\arcsec       \\
Mars    & 850~$\mu$m & 0.23   & 297.1       & 15.3\arcsec      \\
CRL2688 & 850~$\mu$m & 0.23   & 259.6       & 15.3\arcsec      \\ \hline
\end{tabular}
\end{center}
Notes: \fno{a}Sky opacity. \fno{b}Conversion factor from V to Jy
beam$^{-1}$ scale. \fno{c}Beam size (HPBW).
\end{table}

\begin{figure*}
\centering
\resizebox{\hsize}{!}{\rotatebox{90}{\includegraphics{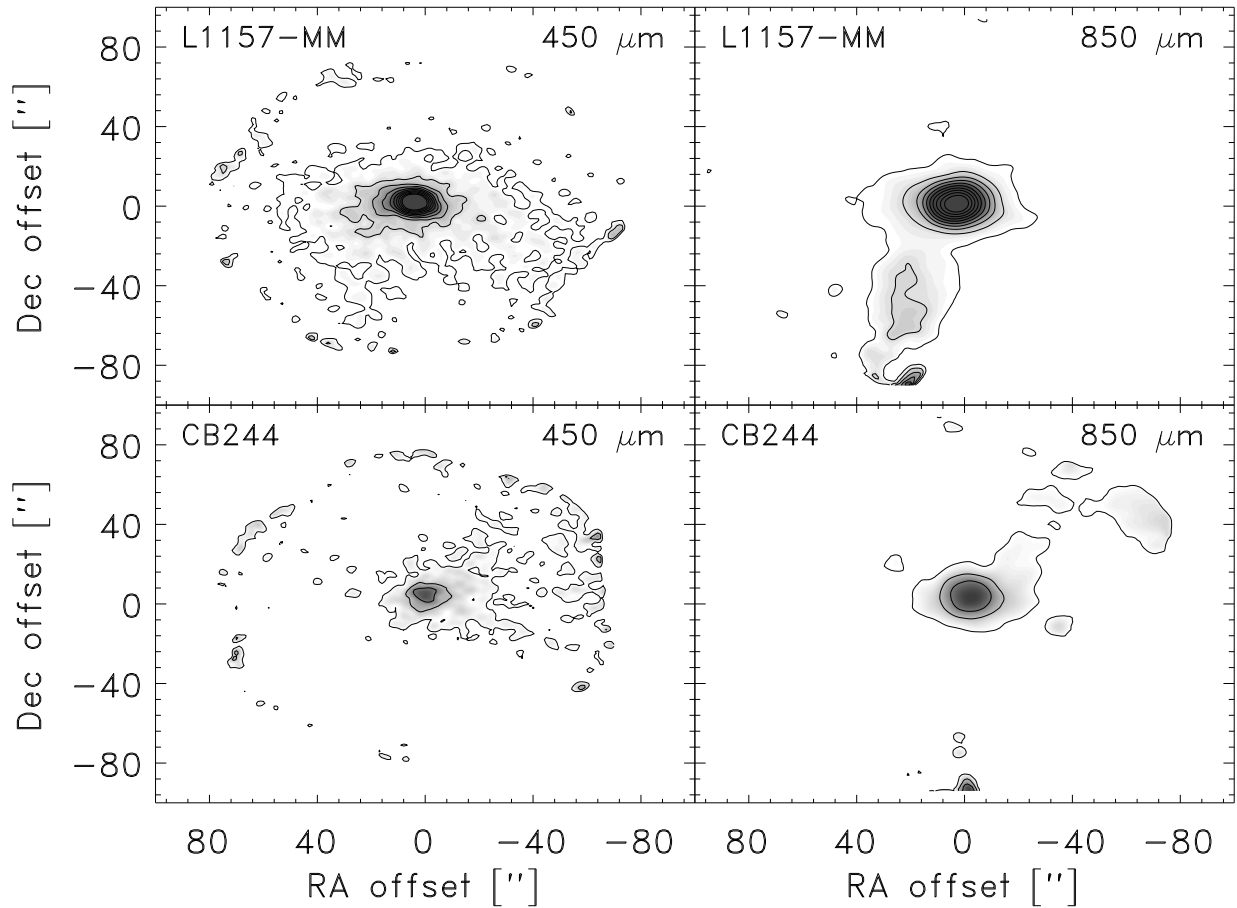}}}
\caption{SCUBA images of \object{CB244} and \object{L1157} at 450 and
850~$\mu$m. The contours indicate the intensity corresponding to
2$\sigma$, 4$\sigma$, etc. with $\sigma$ being the RMS noise given for
each source and wavelength in
Table~\ref{newscuba}.}\label{scubaimages}
\end{figure*}

The derived parameters for \object{L1157} and \object{CB244} are given
in Table~\ref{newscuba}. Images of the two sources at the two SCUBA
wavelengths are presented in Fig.~\ref{scubaimages}. As seen from the
figure, both sources are quite circular with only a small degree of
extended emission. Comparison with the 850~$\mu$m data of
\cite{shirley00} for the 40\arcsec\ aperture shows that the fluxes
agree well within the 20\% uncertainty assumed for the calibration.
\begin{table}
\caption{Results for the \object{CB244} and \object{L1157} submillimeter emission.}\label{newscuba}
\begin{center}
\begin{tabular}{lllll}\hline\hline
 & \multicolumn{2}{c}{\object{CB244}} & \multicolumn{2}{c}{\object{L1157}} \\  
 & 450~$\mu$m & 850~$\mu$m & 450~$\mu$m & 850~$\mu$m \\ \hline
$F_{\rm peak}$\fno{a}  & 3.14 & 0.591 & 8.62 & 1.72  \\
$F_{\rm I,40''}$\fno{b}    & 14.2 & 1.28  & 22.2 & 2.74  \\
$F_{\rm I,120''}$\fno{b}   & 61.4 & 3.11  & 69.2 & 5.87  \\
$F_{\rm noise}$\fno{a} & 0.43 & 0.087 & 0.29 & 0.074 \\ \hline
\end{tabular}
\end{center}
Notes: \fno{a}Peak flux and RMS noise in
Jy beam$^{-1}$. \fno{b}Integrated flux in 40\arcsec\ and 120\arcsec\
apertures respectively in Jy.
\end{table}

\subsection{Line data}
CO line data were obtained with the JCMT in May and August 2001, the
Instituto de Radio Astronomica Milimetrica (IRAM) 30m telescope in
November 2001 and the Onsala Space Observatory 20m telescope in
March 2002, complementing data from the JCMT archive. Our own
observations were performed in beam switching mode using a switch of
180\arcsec\ in declination - except for the sources in NGC1333, where
position switching towards an emission free reference position was
used. A more detailed description of the JCMT and the heterodyne
receivers can be found on the JCMT homepage\footnote{{\tt
http://www.jach.hawaii.edu/JACpublic/JCMT/}}. Where archive data were
available for one line from several different projects, the data
belonging to each observing program were reduced individually and the
results compared giving an estimate of the calibration uncertainty of
the data of 20\%. The integrated line intensities were found by
fitting gaussians to the main line. In some cases outflow or secondary
components were apparent in the line profiles leading to two gaussian
fits. For the C$^{17}$O $J=1-0$ and $J=2-1$ lines, the hyperfine
splitting were apparent, giving rise to two separate lines for the
$J=1-0$ transition separated by about 5 km~s$^{-1}$, while the $J=2-1$
main hyperfine lines are split by less (0.5 km~s$^{-1}$) giving rise
in some cases to line asymmetries. In these cases the quoted line
intensities are the total intensity including all hyperfine lines.

The integrated line data were brought from the antenna temperature
scale $T_{\rm A}^\ast$ to the main-beam brightness scale $T_{\rm mb}$
by dividing by the main-beam brightness efficiency $\eta_{\rm mb}$
taken to be 0.69 for data obtained using the JCMT A band receivers
(210-270 GHz; the $J=2-1$ transitions) and 0.59-0.63 for respectively
the old B3i (before December 1996) and new B3 receivers (330-370 GHz;
the $J=3-2$ transitions). For the IRAM 30m observations beam
efficiencies $B_{\rm eff}$ of 0.74 and 0.54 and forward efficiencies
$F_{\rm eff}$ of 0.95 and 0.91 were adopted for respectively the
C$^{17}$O $J=1-0$ and $J=2-1$ lines, which corresponds to main-beam
brightness efficiencies ($\eta_{\rm mb}=B_{\rm eff}/F_{\rm eff}$) of
respectively 0.78 and 0.59. For the Onsala 20m telescope $\eta_{\rm
mb}=0.43$ was adopted for the C$^{18}$O and C$^{17}$O $J=1-0$
lines. The relevant beam sizes for the JCMT are 21\arcsec\ and
14\arcsec\ at respectively 220 and 330 GHz, for the IRAM 30m,
22\arcsec\ and 11\arcsec\ at respectively 112 and 224 GHz and for the
Onsala 20m, 33\arcsec.  The velocity resolution ranged from 0.1--0.3
km~s$^{-1}$ for the JCMT data and were 0.05 and 0.1 km~s$^{-1}$ for
the observations of respectively the C$^{17}$O $J=1-0$ and $J=2-1$
transitions at the IRAM 30m. The line properties are summarized
later in Sect.~\ref{gasmodelling}.

\section{Continuum modelling}\label{modelling}
\subsection{Input}
To model the physical properties of the envelopes around these sources
the 1D radiative transfer code DUSTY \citep{dusty} was
used.\footnote{DUSTY is publically available from the homepage at {\tt
http://www.pa.uky.edu/$\sim$moshe/dusty/}} The dust grain opacities
from \cite{ossenkopf94} corresponding to coagulated dust grains with
thin ice mantles at a density of $n_{\rm H_2} \sim 10^6\, {\rm
cm}^{-3}$ were adopted. These were found by \cite{vandertak99} to be
the only dust opacities that could reproduce the ``standard''
dust-to-gas mass ratio of 1:100 by comparison to C$^{17}$O
measurements for warm high-mass YSOs where CO is not depleted.

Using a power-law to describe the density leaves five parameters to
fit as summarized in Table~\ref{dustypar}. Not all five parameters are
independent, however: the temperature at the inner boundary, $T_1$,
determines the inner radius of the envelope, $r_1$, through the
luminosity of the source. If the outer radius of the envelope $r_2$ is
expected to be constant, $Y=r_2/r_1$ will depend on the value of
$r_1$, i.e., $T_1$. The results are, however, not expected to depend on
$r_1$ if it is chosen small enough, since the beam size does not
resolve the inner parts anyway. Therefore $T_1$ is simply set to
250~K, a reasonable temperature considering the chemistry observed
towards these sources. Also the temperature of the central star has to
be fixed: a temperature of 5000 K is chosen. This temperature is of
course mostly unknown for the embedded sources, but due to the optical
thickness of the envelope most of the radiation from the central star
is anyway reprocessed by the dust and thus the temperature of the
central star does not play a critical role, e.g., in the resulting
SED.

Although these parameters may not seem the most straightforward
choice, one of the advantages of DUSTY is the scale-free nature
allowing the user to run a large sample of models and then compare a
number of YSOs to these models just by scaling with distance and
luminosity as discussed in \cite{dusty}.
\begin{table}
\caption{Parameters for the DUSTY 1D radiative transfer modelling of
the protostellar envelopes.}\label{dustypar}
\begin{center}
\begin{tabular}{ll}\hline\hline
Param. & Description \\ \hline
\multicolumn{2}{l}{Modelled:} \\
$Y$ & Ratio of the outer ($r_2$) to inner ($r_1$) radius\\
$\tau_{100}$ & Optical depth at 100 $\mu$m \\
$\alpha$ & Density power-law exponent \\ \hline
\multicolumn{2}{l}{Fixed:} \\
$T_1$ & Temperature at inner boundary (250 K) \\
$T_\star$ & Temperature of star (5000 K) \\ \hline
\multicolumn{2}{l}{Literature:} \\ 
$d$ & Distance \\
$L_{\rm bol}$ & Luminosity \\ \hline
\end{tabular}
\end{center}
\end{table}

\subsection{Output}
\begin{figure}
\resizebox{\hsize}{!}{\rotatebox{90}{\includegraphics{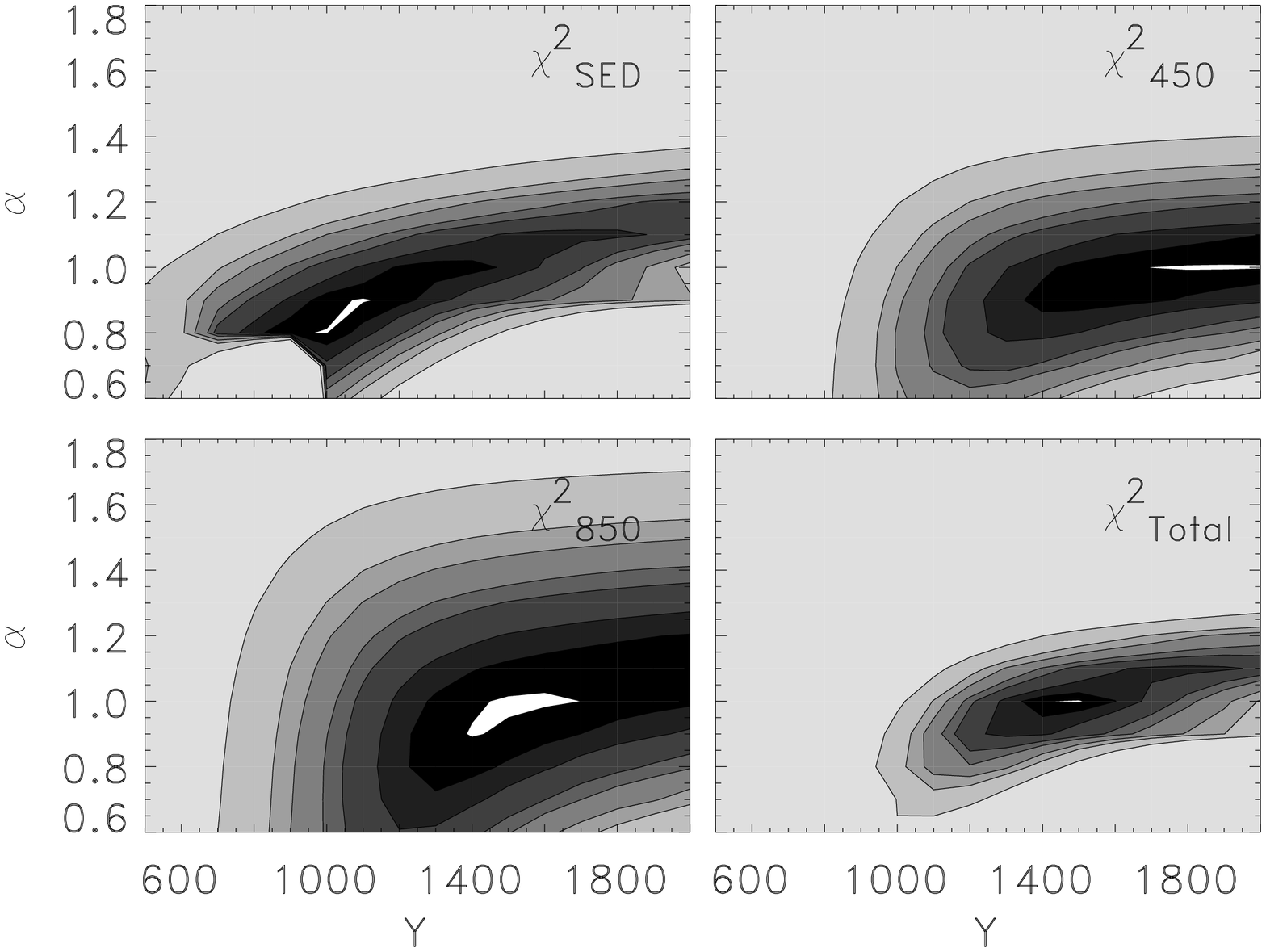}}}
\resizebox{\hsize}{!}{\rotatebox{90}{\includegraphics{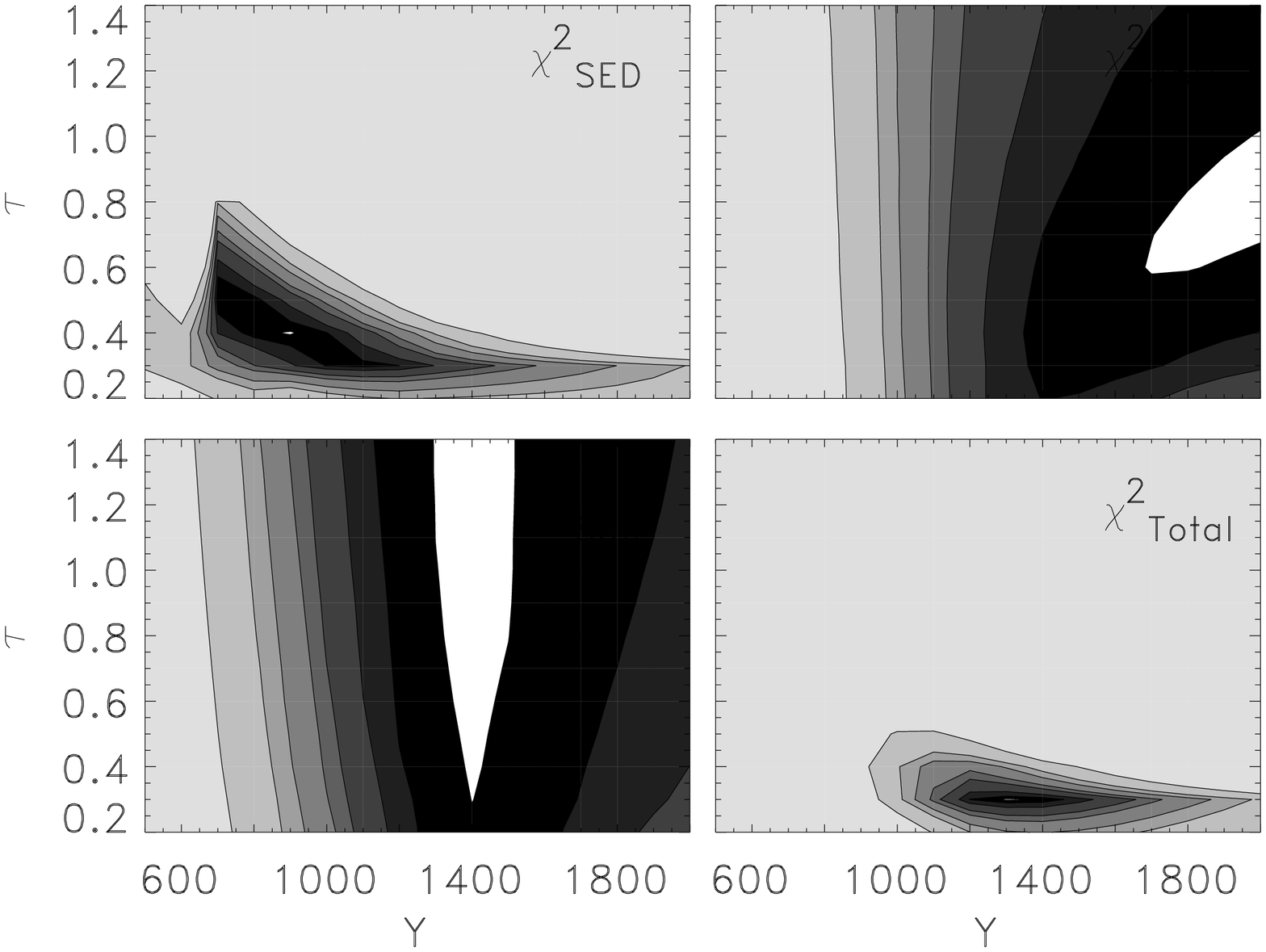}}}
\resizebox{\hsize}{!}{\rotatebox{90}{\includegraphics{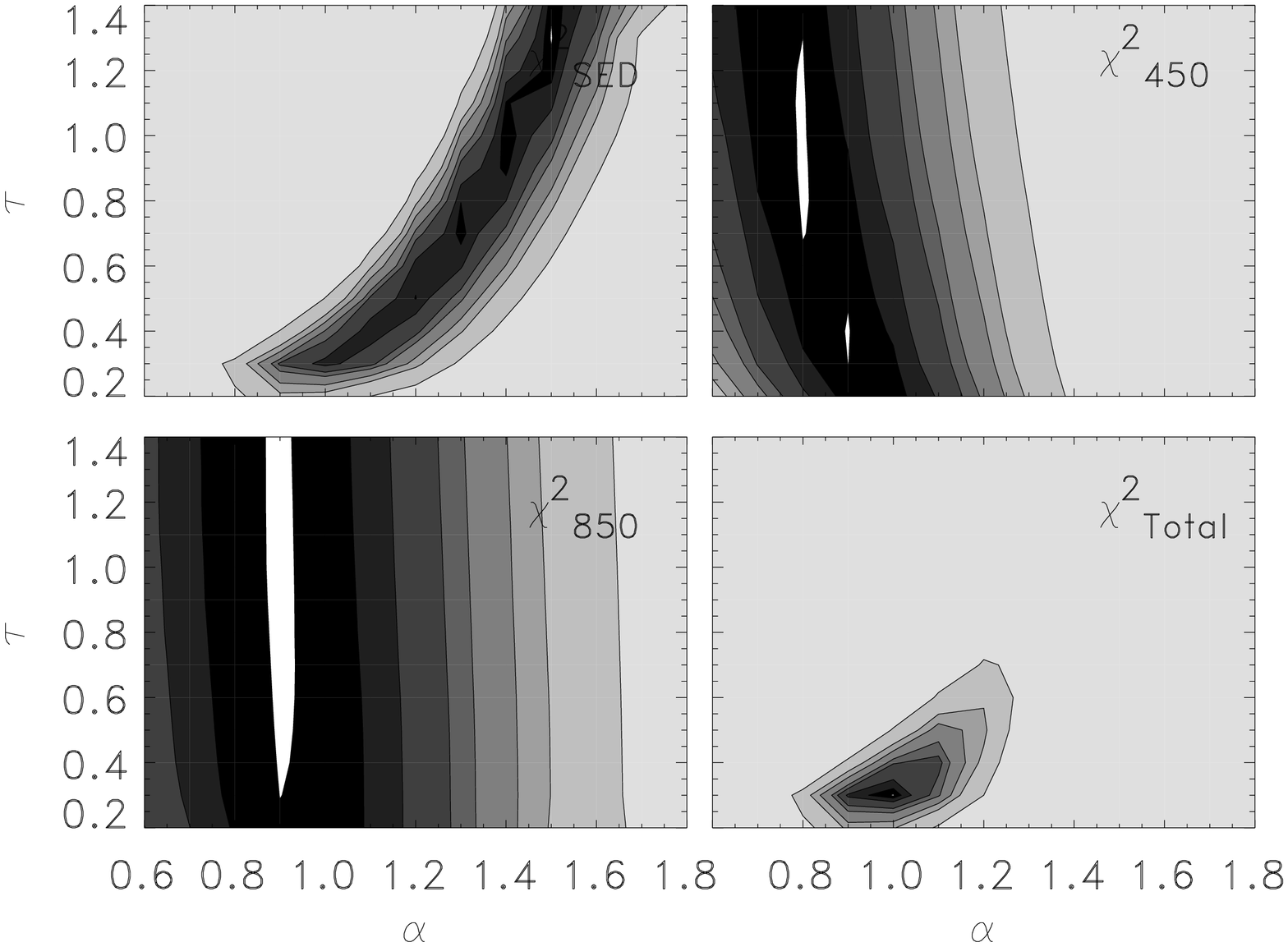}}}
\caption{$\chi^2$ contour plots for the modelling of \object{L483}-mm.
In the four upper panels $\tau_{100}$ is fixed at 0.2, in the 4
middle panels $\alpha$ is fixed at 0.9 and while in the 4 lower panels
$Y$ is fixed at 1400. The solid (dark) contours indicate the
confidence limits corresponding to 1$\sigma$, 2$\sigma$
etc.}\label{chisqr}
\end{figure}

\begin{figure}
\resizebox{\hsize}{!}{\includegraphics{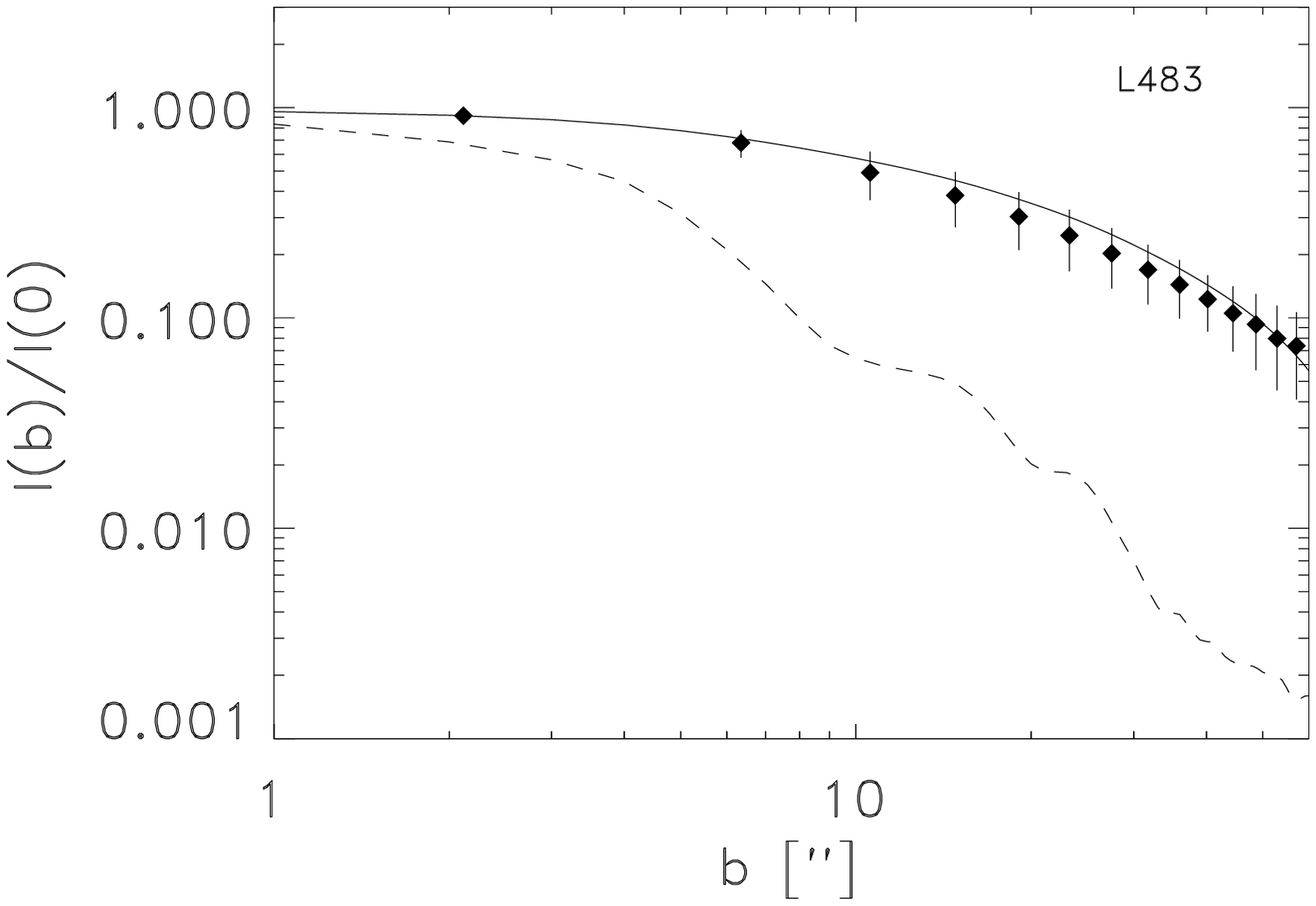}}
\resizebox{\hsize}{!}{\includegraphics{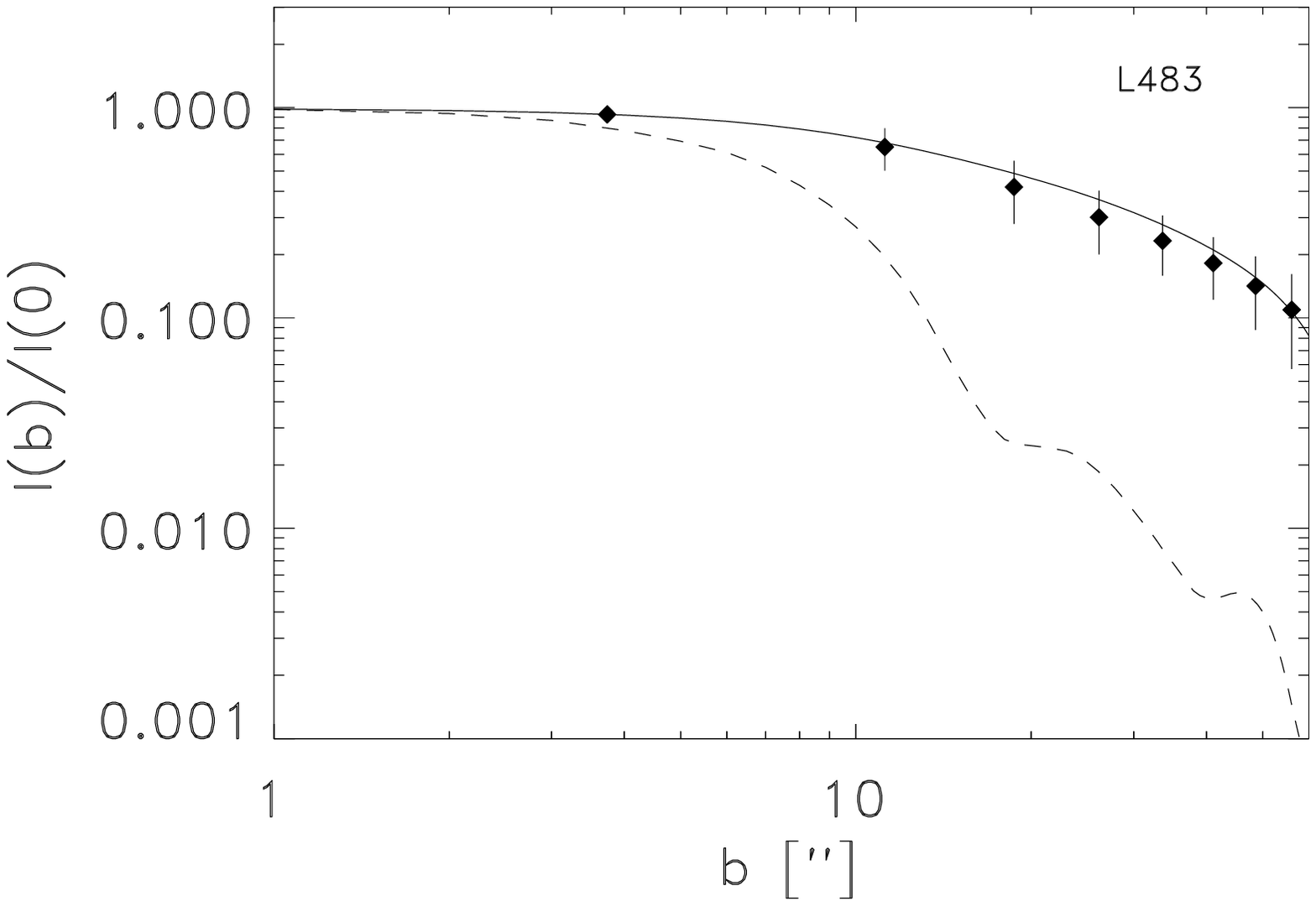}}
\caption{The observed brightness profile for \object{L483} at
450~$\mu$m (upper panel) and 850~$\mu$m (lower panel) with the
best-fit models overplotted (full line). The dashed line indicates the
beam profile used in the modelling.}\label{l483bp}
\end{figure}

\begin{figure}
\resizebox{\hsize}{!}{\includegraphics{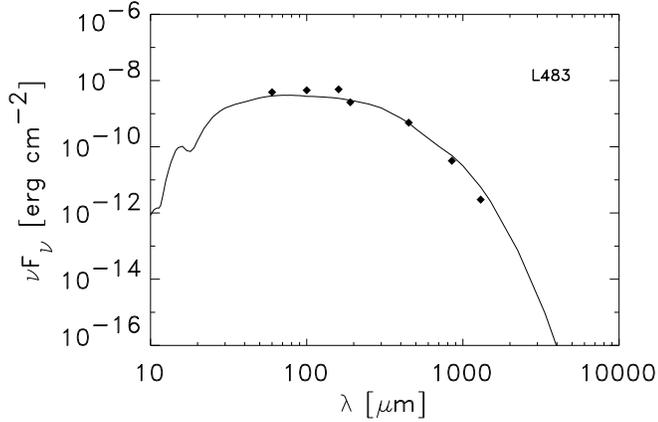}}
\caption{The spectral energy distribution of \object{L483}: the points indicate
the data, the line the best-fit model.}\label{l483sed}
\end{figure}

DUSTY provides fluxes at various wavelengths and brightness profiles
for the sources, which are compared to the SCUBA data and flux
measurements. Given the grid of models the best fit model can then be
determined by calculating the $\chi^2$-statistics for the SED and
brightness profiles at 450 and 850~$\mu$m ($\chi^2_{\rm SED}$,
$\chi^2_{450}$ and $\chi^2_{850}$ respectively). In order to fully
simulate the observations, the modelled brightness profiles are
convolved with the exact beam as obtained from planet
observations. Strictly speaking, the outer parts of the brightness
profile also depend on the chopping of the telescope. The chopping
along one axis does by nature not obey the spherical symmetry, so
simulation of the chopping and comparing this with one dimensional
modelling will not reflect the observations. Therefore in this
analysis only the inner 60\arcsec\ of the brightness profiles are
considered, which (1) should be less sensitive to the typical
120\arcsec\ chop and (2) is typically above the background
emission. For the flux measurements a relative uncertainty of 20\% was
assumed irrespective of what was given in the original reference,
since some authors tend to give only statistical errors and do not
include calibration or systematic errors. By assigning a relative
uncertainty of 20\% to all measurements each point is weighted equal
but more weight is given to a given part of the SED if several
independent measurements exist around a certain wavelength. Contour plots of the derived $\chi^2$ values for L483-mm are presented
in Fig.~\ref{chisqr}, while the actual fits to the brightness profiles
and the SED for this source are shown in Fig.~\ref{l483bp} and
\ref{l483sed}.

In determining the best fit model each of the calculated $\chi^2$
values are considered individually. The total $\chi^2$ obtained by
adding the $\chi^2_{\rm SED}$, $\chi^2_{850}$ and $\chi^2_{450}$ does
not make sense in a strictly statistical way, since the observations
going into these cases are not 100\% independent. Another reason for
not combining the values of $\chi^2_{\rm SED}$, $\chi^2_{450}$ and
$\chi^2_{850}$ into one total $\chi^2$ is that the parameters
constrained by the SED and brightness profiles are different. For
example the brightness profiles provide good constraints on $\alpha$
as seen in a 2D contour ($Y,\alpha$) plot of, e.g., $\chi^2_{450}$ in
Fig.~\ref{chisqr}, while these do not depend critically on the value
of $\tau_{100}$. The most characteristic feature of the
$\chi^2$-values for the SEDs on the other hand is the band of possible
models in contour plots for ($\alpha,\tau_{100}$), giving an almost
one-to-one correspondence for a best fit $\tau_{100}$ for each value
$\alpha$.

These features are actually easily understood: $\chi^2_{450}$ and
$\chi^2_{850}$ are the normalized profiles and should thus not depend
directly on the value of $\tau_{100}$. On the other hand since the
peaks of the SEDs are typically found at wavelengths longer than 100
$\mu$m, increasing $\tau_{100}$ and thus the flux at this wavelength,
require the best-fit SED to shift towards 100~$\mu$m, i.e., with less
material in the outer cool parts of the envelope, which can be
obtained by a steeper value of $\alpha$. The value of $Y$ is less well
constrained, mainly because of its relation to the temperature at the
inner radius (and through that the luminosity of the central
source). As illustrated in Fig.~\ref{chisqr}, $Y$ is constrained
within a factor 1.5-2.0 at the $2 \sigma$ level, but the question is
how physical the outer boundary of the envelope is: is a sharp outer
boundary expected or rather a soft transition as the density and
temperature in the envelope reaches that of the surrounding molecular
cloud? In the first case, a clear drop of the observed brightness
profile should be seen compared with a model with a (sufficiently)
large value of $Y$, e.g., corresponding to an outer temperature of
5--10 K, less than the temperature of a typical molecular cloud. In
the other case, however, such a model will be able to trace the
brightness profile all the way down to the noise limit. The
modelling of the CO lines (see Sect.~\ref{gasmodelling}) indicates
that significant ambient cloud material is present towards most
sources, so the transition from the isolated protostar to the parental
cloud is likely to be more complex than described by a single
power-law.

The features of the values of $\chi^2_{\rm SED}$, $\chi^2_{450}$ and
$\chi^2_{850}$ provide an obvious strategy for selecting the best fit
models: first the best fit value of $\alpha$ is selected on the basis
of the brightness profiles and second the corresponding value of
$\tau_{100}$ is selected from the $\chi^2_{\rm SED}$ contour
plots. For a few sources there is not 100\% overlap between the 450
and 850~$\mu$m brightness profiles and it is not clear which
brightness profile is better. The beam at 850~$\mu$m is significantly
larger than that at 450~$\mu$m (15\arcsec\ vs. 9\arcsec) and even
though the beam is taken into account explicitly, the sensitivity of
the 850~$\mu$m data to variations in the density profile must be lower
as is seen from Fig.~\ref{chisqr}. On the other hand, the 450~$\mu$m
data generally suffer from higher noise, so especially the weak
emission from the envelope, which provides the better constraints on
the outer parts of the envelope and thus the power-law exponent, will
be more doubtful at this wavelength. The power-law slopes found from
modelling the two brightness profiles agree, however, within the
uncertainties ($\alpha \sim \pm 0.2$).

\subsection{Results}
\begin{figure*}
\centering
\resizebox{\hsize}{!}{\rotatebox{90}{\includegraphics{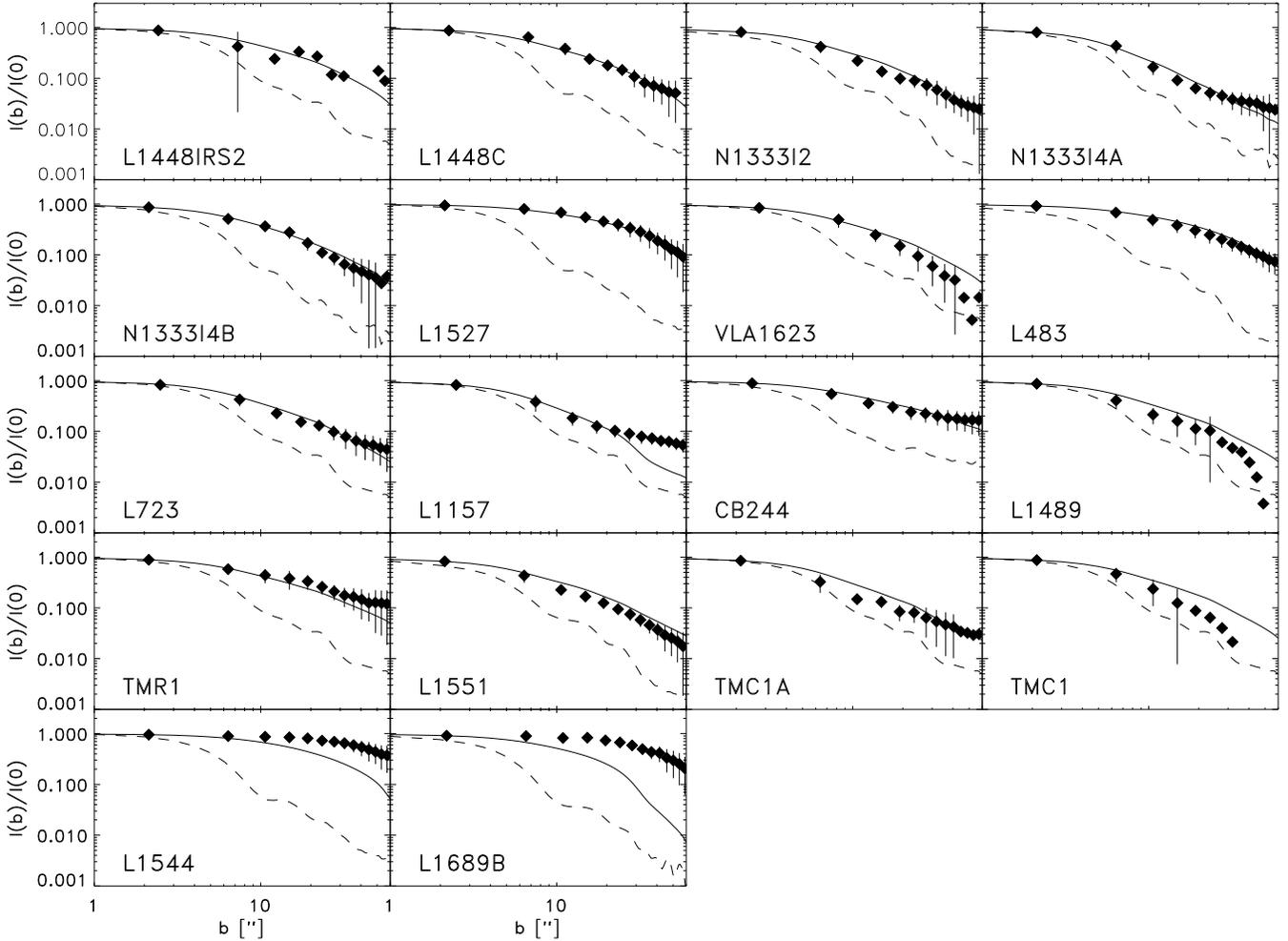}}}
\caption{Composite showing the brightness profiles of the sources
overplotted with the best fit model at 450~$\mu$m. The dashed line
indicate the beam profile.}\label{bright450sum}
\end{figure*}

\begin{figure*}
\centering
\resizebox{\hsize}{!}{\rotatebox{90}{\includegraphics{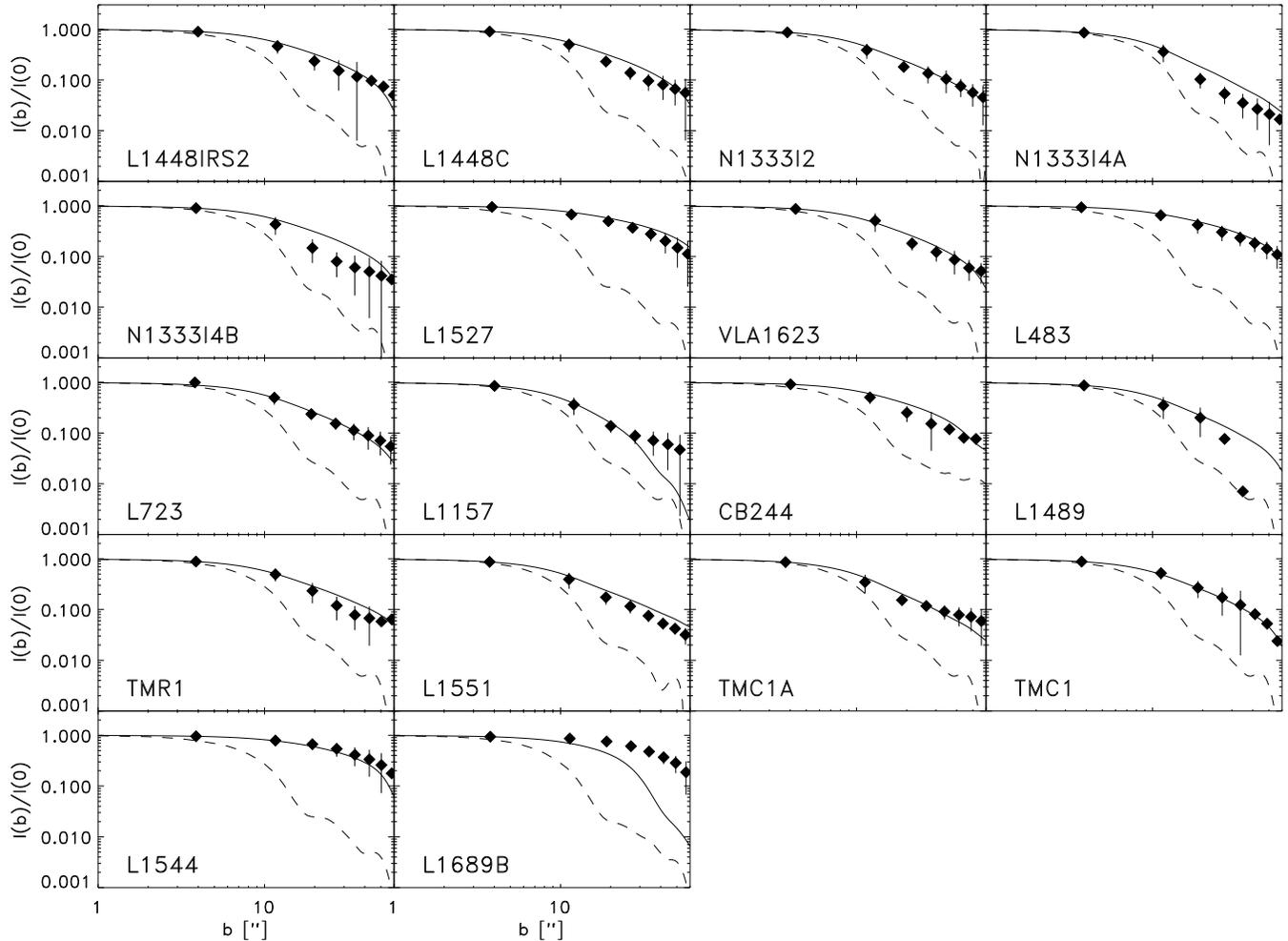}}}
\caption{As in Fig.~\ref{bright450sum} but for the 850~$\mu$m
data.}\label{bright850sum}
\end{figure*}

\begin{figure*}
\centering
\resizebox{\hsize}{!}{\rotatebox{90}{\includegraphics{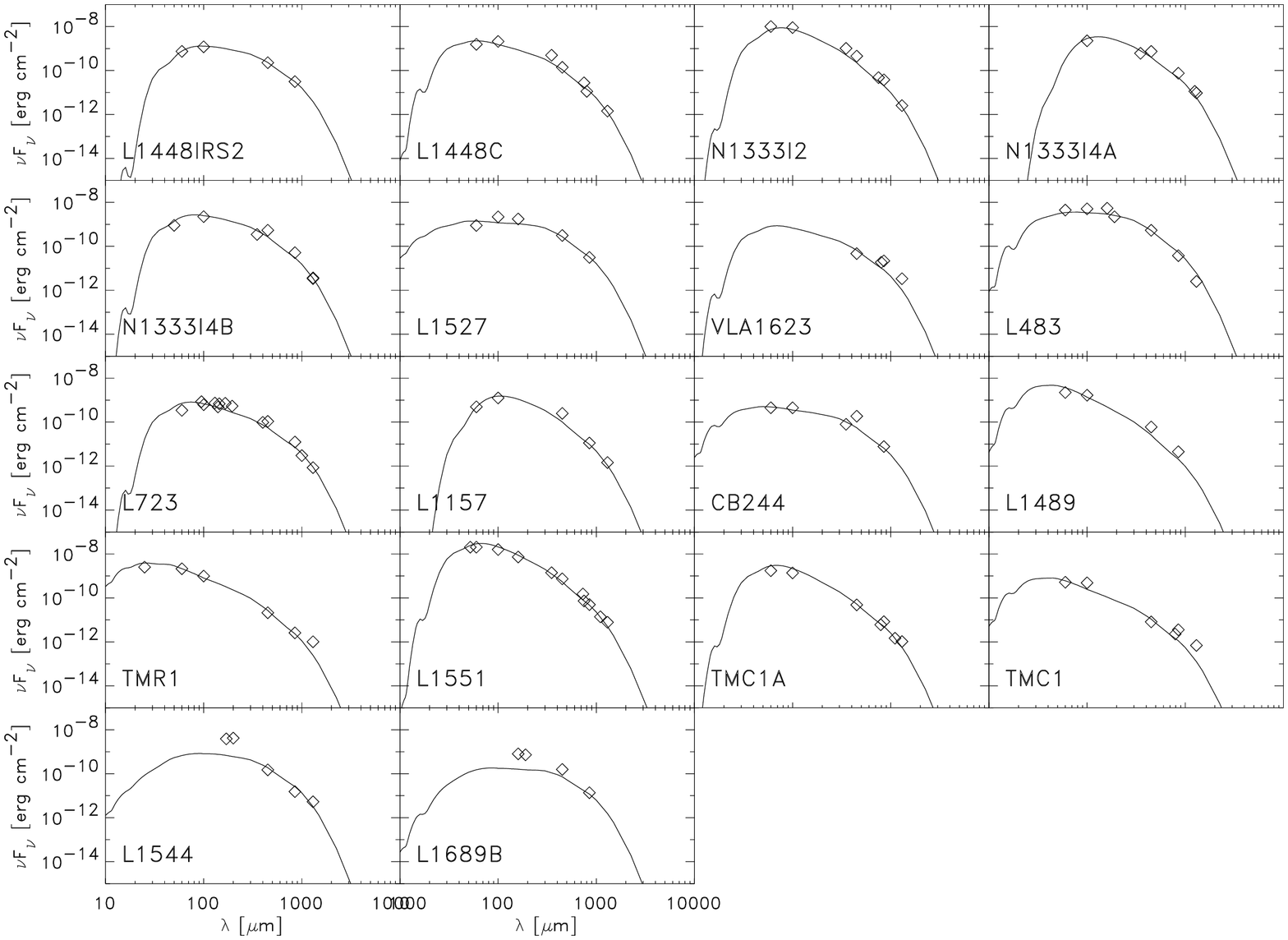}}}
\caption{Composite showing the SEDs of each source overplotted with
the SED from the best fit model. The individual SEDs are based on
literature searches, with the main references being \cite{shirley00}
(class 0 objects and pre-stellar cores), \cite{chandler00}
(NGC1333-I2), \cite{sandell91} (NGC1333-I4A,B; 350 and 800 $\mu$m),
IRAS Faint Source Catalog (\object{TMR1}, \object{L1489}) and 1.3 mm
fluxes from \cite{motte01} for sources included in their
sample.}\label{sedsum}
\end{figure*}

In Table~\ref{dustyfitpar}, the fitted values of the three parameters
for each source are presented and in Table~\ref{dustyphyspar} the
physical parameters obtained by scaling according to source distance
and luminosities are given. As obvious from the $\chi^2$ plots in
Fig.~\ref{chisqr}, these parameters have some associated
uncertainties. The value of $\alpha$ is determined typically within
$\pm 0.2$ leading to a similar uncertainty in $\tau_{100}$ of $\pm
0.2$. The minimum value of $Y$ gives a corresponding minimum value of
the outer radius. As discussed above, increasing $Y$ only corresponds
to adding more material after the outer boundary, so if $Y$ is large
enough to encompass the point where the temperature $T$ reaches 10~K,
the radius corresponding to this temperature can be used as a
characteristic size of the envelope. The region of the envelopes
within this radius corresponds to the inner 40-50\arcsec\ of the
brightness profiles for all the sources. If one would increase the
radius further, the typical 120\arcsec\ chop throw for SCUBA should be
taken into account when comparing the brightness profiles from the
models with the observations. This could lead to flatter density
profiles with $\alpha$ decreased by $\sim 0.2$
\citep[e.g.][]{motte01}. Although the models presented here do not
extend that far, one should still be aware of the possibility that
emission is picked up at the reference position, which would lead to
an overestimate of the steepness of the density distribution. The
fitted brightness profiles and SEDs for all sources are presented in
Fig.~\ref{bright450sum}, \ref{bright850sum} and \ref{sedsum}.
\begin{table}
\caption{Best fit parameters from DUSTY modelling.} \label{dustyfitpar}
\begin{center}
\begin{tabular}{llll}\hline\hline
Source & $Y$\fno{a} & $\alpha$ & $\tau_{100}$  \\ \hline
\object{L1448-I2} & 1800 & 1.2 & 1.1\\
\object{L1448-C} & 1600 & 1.4 & 0.5\\
\object{N1333-I2} &  900 & 1.8 & 1.6\\
\object{N1333-I4A} & 1000 & 1.8 & 6.5\\
\object{N1333-I4B} & 1400 & 1.3 & 0.9\\
\object{L1527} & 2500 & 0.6 & 0.1\\
\object{VLA1623} & 2400 & 1.4 & 0.7\\
\object{L483} & 1400 & 0.9 & 0.3\\
\object{L723} & 2500 & 1.5 & 1.0\\
\object{L1157} &  600 & 1.7 & 3.4\\
\object{CB244} & 2600 & 1.1 & 0.2\\
\object{L1489} & 1200 & 1.8 & 0.3\\
\object{TMR1} & 2000 & 1.6 & 0.1\\
\object{L1551-I5} & 1000 & 1.8 & 1.1\\
\object{TMC1A} & 1700 & 1.9 & 1.3\\
\object{TMC1} & 2900 & 1.6 & 0.2\\
\object{L1544} & 2800 & 0.1 & 0.1\\
\object{L1689B} & 3000 & 0.1 & 0.2\\
\hline
\end{tabular}
\end{center}
Notes: \fno{a}Value corresponding to $r_2/r_1$ in Table~\ref{dustyphyspar}.
\end{table}

\begin{table*}
\caption{Result of DUSTY modelling - derived physical parameters.}  \label{dustyphyspar}
\begin{center}
\begin{tabular}{lllllllll}\hline\hline 
Source & $r_1$ & $r_2$ & $r_{\rm 10 K}$ & $N_{\rm H_2,10 K}$ & $M_{\rm 10 K}$ & $n(r_1)$ & $n_{1000 {\rm AU}}$ & $n_{10 {\rm K}}$ \\
 & (AU) & (AU) & (AU) & (cm$^{-2}$) & (M$_\odot$) & (cm$^{-3}$) & (cm$^{-3}$) & (cm$^{-3}$) \\ \hline 
\object{L1448-I2} &  7.4 &  $  1.3\times 10^{ 4} $ &  $  4.5\times 10^{ 3} $ &  $  3.5\times 10^{23} $ &  1.5 &  $  8.8\times 10^{ 8} $ &  $  2.5\times 10^{ 6} $ &  $  4.0\times 10^{ 5} $ \\
\object{L1448-C} &  9.0 &  $  1.5\times 10^{ 4} $ &  $  8.1\times 10^{ 3} $ &  $  1.7\times 10^{23} $ & 0.93 &  $  5.4\times 10^{ 8} $ &  $  7.5\times 10^{ 5} $ &  $  4.0\times 10^{ 4} $ \\
\object{N1333-I2} & 23.4 &  $  2.1\times 10^{ 4} $ &  $  1.2\times 10^{ 4} $ &  $  5.5\times 10^{23} $ &  1.7 &  $  1.3\times 10^{ 9} $ &  $  1.5\times 10^{ 6} $ &  $  1.7\times 10^{ 4} $ \\
\object{N1333-I4A} & 23.9 &  $  2.4\times 10^{ 4} $ &  $  4.7\times 10^{ 3} $ &  $  2.2\times 10^{24} $ &  2.3 &  $  5.0\times 10^{ 9} $ &  $  6.3\times 10^{ 6} $ &  $  3.8\times 10^{ 5} $ \\
\object{N1333-I4B} & 10.6 &  $  1.5\times 10^{ 4} $ &  $  7.0\times 10^{ 3} $ &  $  3.0\times 10^{23} $ &  2.0 &  $  6.7\times 10^{ 8} $ &  $  1.8\times 10^{ 6} $ &  $  1.4\times 10^{ 5} $ \\
\object{L1527} &  4.2 &  $  1.1\times 10^{ 4} $ &  $  6.3\times 10^{ 3} $ &  $  2.8\times 10^{22} $ & 0.91 &  $  9.9\times 10^{ 6} $ &  $  3.8\times 10^{ 5} $ &  $  1.2\times 10^{ 5} $ \\
\object{VLA1623} &  4.3 &  $  1.0\times 10^{ 4} $ &  $  3.3\times 10^{ 3} $ &  $  2.3\times 10^{23} $ & 0.22 &  $  1.6\times 10^{ 9} $ &  $  7.7\times 10^{ 5} $ &  $  1.5\times 10^{ 5} $ \\
\object{L483} & 21.5 &  $  3.2\times 10^{ 4} $ &  $  7.8\times 10^{ 3} $ &  $  9.3\times 10^{23} $ &  1.1 &  $  2.7\times 10^{ 9} $ &  $  1.7\times 10^{ 6} $ &  $  3.4\times 10^{ 4} $ \\
\object{L723} &  8.2 &  $  2.1\times 10^{ 4} $ &  $  5.4\times 10^{ 3} $ &  $  3.4\times 10^{23} $ & 0.62 &  $  1.4\times 10^{ 9} $ &  $  1.1\times 10^{ 6} $ &  $  8.6\times 10^{ 4} $ \\
\object{L1157} & 17.9 &  $  1.1\times 10^{ 4} $ &  $  5.4\times 10^{ 3} $ &  $  1.2\times 10^{24} $ &  1.6 &  $  3.1\times 10^{ 9} $ &  $  3.3\times 10^{ 6} $ &  $  1.9\times 10^{ 5} $ \\
\object{CB244} &  3.3 &  $  8.7\times 10^{ 3} $ &  $  4.3\times 10^{ 3} $ &  $  6.5\times 10^{22} $ & 0.28 &  $  2.5\times 10^{ 8} $ &  $  4.8\times 10^{ 5} $ &  $  9.7\times 10^{ 4} $ \\
\object{L1489} &  7.8 &  $  9.4\times 10^{ 3} $ &  $  9.2\times 10^{ 3} $ &  $  1.0\times 10^{23} $ & 0.097 &  $  7.1\times 10^{ 8} $ &  $  1.2\times 10^{ 5} $ &  $  2.1\times 10^{ 3} $ \\
\object{TMR1} &  6.7 &  $  1.3\times 10^{ 4} $ &  $  1.2\times 10^{ 4} $ &  $  3.5\times 10^{22} $ & 0.12 &  $  2.1\times 10^{ 8} $ &  $  6.9\times 10^{ 4} $ &  $  1.3\times 10^{ 3} $ \\
\object{L1551-I5} & 24.8 &  $  2.5\times 10^{ 4} $ &  $  1.6\times 10^{ 4} $ &  $  3.8\times 10^{23} $ &  1.7 &  $  8.2\times 10^{ 8} $ &  $  1.1\times 10^{ 6} $ &  $  7.2\times 10^{ 3} $ \\
\object{TMC1A} &  8.4 &  $  1.4\times 10^{ 4} $ &  $  4.8\times 10^{ 3} $ &  $  4.5\times 10^{23} $ & 0.13 &  $  3.2\times 10^{ 9} $ &  $  3.7\times 10^{ 5} $ &  $  1.8\times 10^{ 4} $ \\
\object{TMC1} &  3.1 &  $  9.0\times 10^{ 3} $ &  $  4.3\times 10^{ 3} $ &  $  6.9\times 10^{22} $ & 0.034 &  $  9.0\times 10^{ 8} $ &  $  8.8\times 10^{ 4} $ &  $  8.5\times 10^{ 3} $ \\
\object{L1544} &  3.0 &  $  8.3\times 10^{ 3} $ &  $  4.0\times 10^{ 3} $ &  $  1.8\times 10^{22} $ & 0.41 &  $  5.5\times 10^{ 5} $ &  $  3.1\times 10^{ 5} $ &  $  2.7\times 10^{ 5} $ \\
\object{L1689B} &  1.3 &  $  4.0\times 10^{ 3} $ &  $  1.5\times 10^{ 3} $ &  $  2.9\times 10^{22} $ & 0.096 &  $  2.3\times 10^{ 6} $ &  $  1.2\times 10^{ 6} $ &  $  1.1\times 10^{ 6} $ \\
\hline
\end{tabular}
\end{center}
\end{table*}


\begin{figure}
\resizebox{\hsize}{!}{\includegraphics{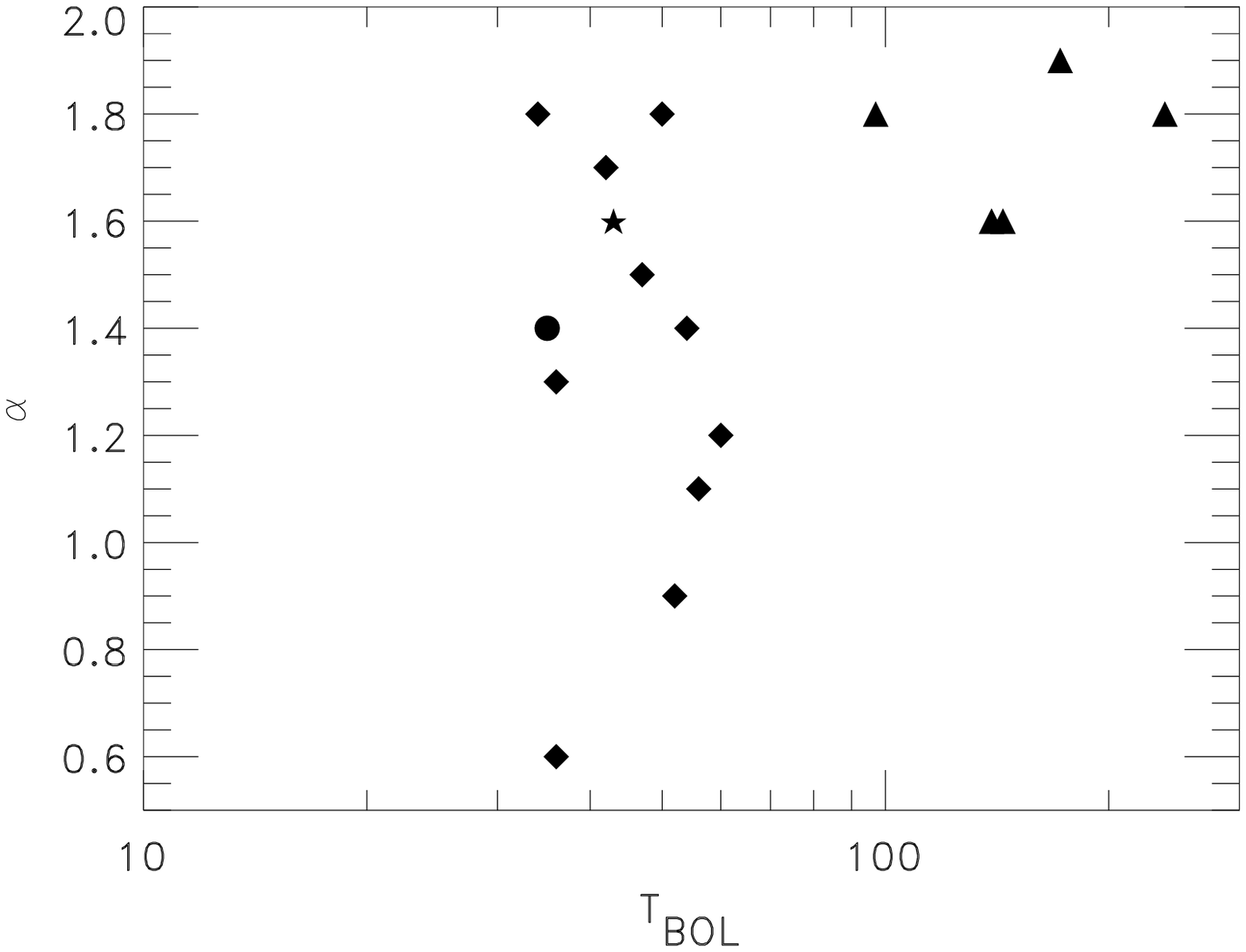}}
\resizebox{\hsize}{!}{\includegraphics{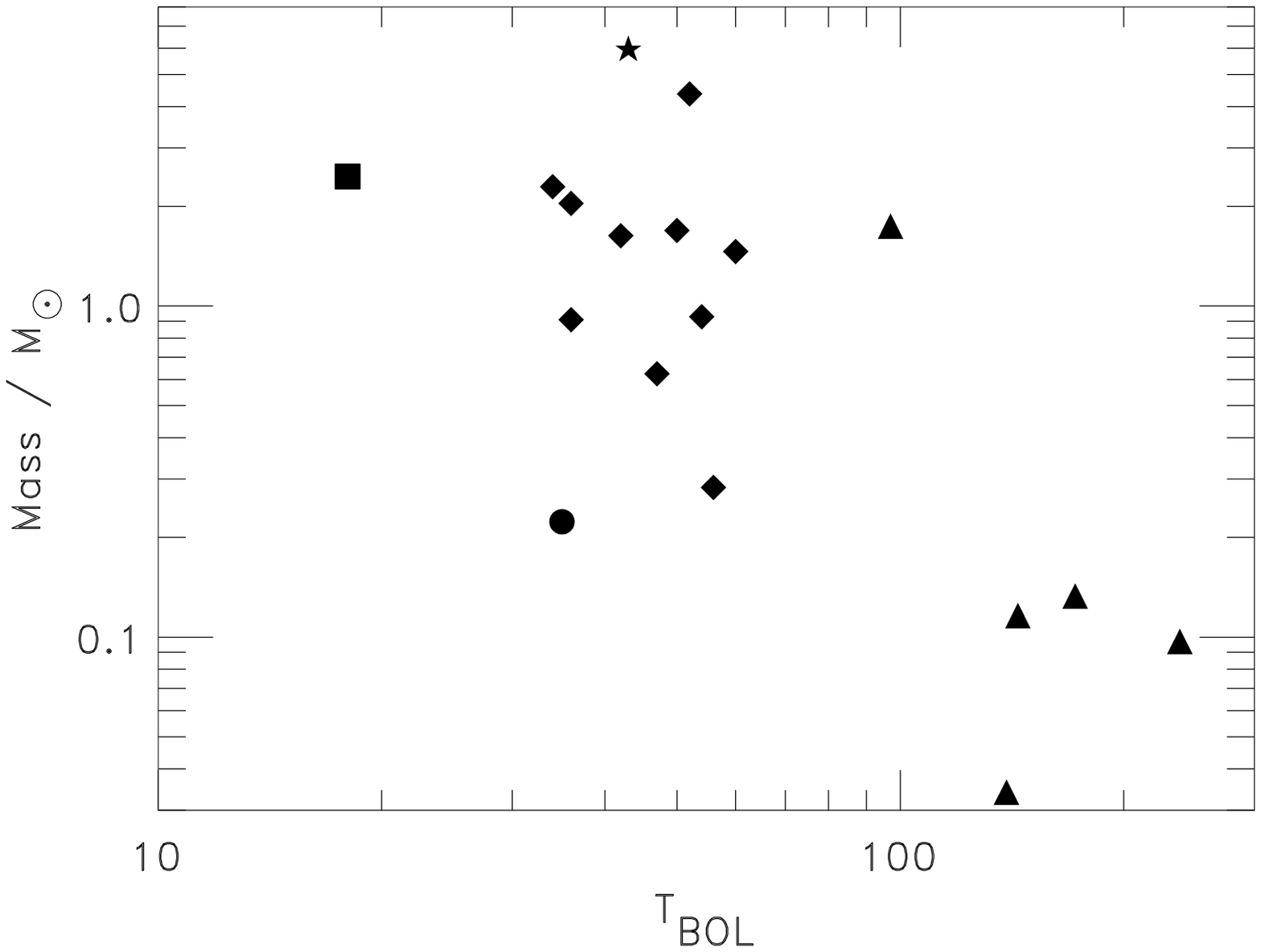}}
\caption{The slope of the density distribution (upper panel) and the
mass within the 10~K radius (lower panel) vs. the bolometric
temperature of the sources. Class 0 objects are marked by
``$\blacklozenge$'', class I objects by ``$\blacktriangle$'' and pre-stellar
cores by ``$\blacksquare$''. VLA1623 and IRAS 16293-2422 have been singled
out with respectively ``{\large $\bullet$}'' and ``$\bigstar$''. The mass of the
pre-stellar cores in the lower panel is the mass of the Bonnor-Ebert
sphere adopted for the line modelling.}\label{tbolfysparam}
\end{figure}

The derived power-law indices are for most sources in agreement with
the predictions from the inside-out collapse model of
$\alpha=1.5-2.0$. A few YSOs (esp. \object{L1527}, \object{L483},
\object{CB244} and \object{L1448-I2}) have density distributions that
are flatter, but as argued in Sect.~\ref{geometry}, this can be
explained by the fact that these sources seem to have significant
departures from spherical symmetry. Apart from these sources, there is
no apparent distinction between the class 0 and class I sources in the
sample as shown in the plot of power-law slope versus bolometric
temperature in the upper panel of Fig.~\ref{tbolfysparam}. There is,
however, a significant difference in the masses derived for these
types of objects, with the class 0 objects having significantly more
massive envelopes (lower panel of Fig.~\ref{tbolfysparam}). This is to
be expected since the bolometric temperature measures the redness (or
coolness) of the SED, i.e., the amount of envelope material. Finally,
there is no dependence on either mass or power-law slope with
distance, which strengthens the validity of the derived parameters.

\subsection{Individual sources}\label{indsources}
For some of the sources the derived results are uncertain for various
reasons, e.g., the interpretation of their surrounding
environment. These cases together with other interesting properties of
the sources are briefly discussed below. The failure of our power-law
approach to describe the pre-stellar cores will be further discussed
in Sect.~\ref{ppcs}.
\begin{description}
\item[N1333-I2,-I4:] A major factor of uncertainty in the
determination of the parameters for the sources associated with the
reflection nebula NGC1333 is the distance of these sources. Values
ranging from 220~pc \citep{cernis90} to 350~pc \citep{herbig83} have
been suggested. The latter determination of the distance assumes that
NGC 1333 (and the associated dark cloud L1450) is part of the Perseus
OB2 association, which recent estimates place at $318\pm27$~pc
\citep{dezeeuw99}. The first value is a more direct estimate
based on the extinction towards the
cloud. \object{N1333-I2} can be modelled using either distance - but
the 10 K radius becomes rather large, i.e., 19000~AU in the case of an
assumed distance of 350 pc, while the distance of 220~pc leads to a
radius of 11000~AU that is more consistent with those of the other
sources.

\item[] N1333-I4 is one of the best-studied low-mass protostellar
systems, both with respect to the molecular content \citep{blake95}
and in interferometric continuum studies \citep{looney00}. On the
largest scales the entire system is seen to be embedded in a single
envelope, but going to progressively smaller scales shows that both
\object{N1333-I4A} and \object{N1333-I4B} are multiple in nature
\citep{looney00}. The small separation between the two sources can
cause problems when interpreting the emission from the envelopes of
each of the sources. On the other hand the small scale binary
components of \object{N1333-I4A} and \object{N1333-I4B} each should be
embedded in common envelopes and can at most introduce a departure
from the spherical symmetry.

\item[L1448-C, -I2:] The L1448 cloud reveals a complex of 4 or more
class 0 objects with \object{L1448-C} (or L1448-mm) and its powerful outflow
together with the binary protostar L1448-N being well studied
\citep[e.g.,][]{barsony98}. Also the recently identified \object{L1448-I2}
\citep{olinger99} shows typical protostellar properties. The dark
cloud L1448 itself is a member of the Perseus molecular cloud complex,
for which we adopt a distance of 220~pc (see above). Note, however, that
\cite{cernis90} mentions the possibility of a distance gradient across
the cloud complex so that larger distances may be appropriate for the
L1448 objects.
\item[VLA 1623:] As the first ``identified'' class 0 object, VLA 1623
is often discussed as a prototype class 0 object. It is, however, not
well suited for discussions of the properties of these objects because
of its location close to a number of submillimeter cores
\citep[e.g.,][]{wilson99}. This makes it hard to extract and model the
properties of this source and might explain why it has been claimed to
have a very shallow density profile of $\rho\propto r^{-0.5}$
\citep{andre93} or a constant density outer envelope
\citep{jayawardhana01}. If the emission in the three quadrants towards
the other submillimeter cores is blocked out when creating the
brightness profiles it is found that VLA 1623 can be modelled with an
almost ``standard'' density profile with $\alpha=1.4$, although with
rather large uncertainties.
\item[L1527:] \object{L1527} is remarkable for its rather flat
envelope profile with $\alpha\approx 0.6$. It was one of the sources
for which \cite{chandler00} estimated the relative contribution of the
disk and envelope to the total flux at 450 and 850~$\mu$m and found
that the envelope contributes by more than 85\%, which justifies our
use of the images at these wavelengths to constrain the envelope. On
the other hand, \cite{hogerheijde97} estimate the disk contribution at
1.1 mm in a 19\arcsec\ beam to be between 30 and 75\% of the continuum
emission ($\approx$ 50\% for \object{L1527}) for a sample of mainly
class I objects, so it is evident that possible disk emission is a
factor of uncertainty in the envelope modelling. Disk emission would
contribute to the fluxes of the innermost points on the brightness
profiles and so lead to a steeper density profile.
\item[L723-mm:] The most characteristic feature about \object{L723}-mm
is the quadrupolar outflow originating in the central source, which
has lead to the suggestion that the central star is a binary
\citep{girart97}.
\item[L483-mm:] \object{L483}-mm is a good example of a central source with an
asymmetry yet providing an excellent fit to the brightness profile
with the simple power-law (see discussion in
Sect.~\ref{geometry}). The source seems to be located in a flattened
filament showing up clearly in the SCUBA maps
\citep[e.g.,][]{shirley00} and integrated NH$_3$ emission
\citep{fuller00}.
\item[L1157-mm:] \object{L1157}-mm is not as well known for the
protostellar source itself as for its bipolar outflow, where a large
enhancement of chemical species like CH$_3$OH, HCN and H$_2$CO is seen
\citep[e.g.,][]{bachiller97}. As a result, the SED of the protostar
itself is rather poorly determined. Our images in Fig.~\ref{newscuba}
show a source slightly extended in the east-west direction and with a
second object showing up south of the source in the 850~$\mu$m
data. Recently \cite{chini01} reported a similar observation of the
source and added that the southern feature is also seen in the 1.3 mm
data in the direction of the CO outflow from \object{L1157},
suggesting an interaction between the outflowing gas and the
circumstellar dust.
\item[CB244:] \object{CB244} is the only protostar of our sample not
included in the table of \cite{andreppiv}. \cite{launhardt97wh} found
that this relatively isolated globule indeed has a high submillimeter
flux, $L_{\rm bol}/L_{\rm sub mm} \geq 2\%$ qualifying it as a class 0
object. It is, however, probably close to the boundary between the class
0 and I stages: \cite{saraceno96} found that it falls in the area of
the class I objects in a $L_{\rm bol}$ vs. $F_{\rm mm}$ diagram.
\item[L1489:] Of the two class I sources in our main sample,
\object{L1489} has recently drawn attention with the suggestion that
the central star is surrounded by a disk-like structure rather than
the ``usual'' envelope for class 0/I objects
\citep{hogerheijde00sandell,hogerheijde01}. \citeauthor{hogerheijde00sandell}
examined the SCUBA images of this source in comparison with the line
emission with the purpose of testing the different models for the
envelope structure - especially the \cite{shu77} infall model. They
found that \object{L1489} could not be fitted in the inside-out
collapse scenario, if both the SCUBA images and spectroscopic data are
modelled simultaneously. Instead they suggested that \object{L1489} is
an object undergoing a transition from the class I to II stages,
revealing a 2000~AU disk, whose velocity structure is revealed through
high resolution HCO$^+$ interferometer data
\citep{hogerheijde01}. That we actually can fit the SCUBA data is
neither proving nor disproving this
result. \citeauthor{hogerheijde00sandell} in fact remark that it is
possible to fit the continuum data alone, but that this would
correspond to an unrealistic high age of this source. The modelling of
\object{L1489} is slightly complicated by a nearby submillimeter
condensation - presumably a pre-stellar core, which has to be blocked
out leading to an increase in the uncertainty for the fitting of the
brightness profile.
\item[TMR1:] This is a more standard class I object showing a bipolar
nebulosity in the infrared corresponding to the outflow cavities of
the envelope \citep{hogerheijde98}.
\end{description}

\section{Discussion and comparison}\label{discuss}
\subsection{Power law or not?}
The first simplistic assumption in the above modelling is (as in other
recent works, e.g., \cite{shirley00, chandler00, motte01}) that the
density distribution can be described by a single power law. This is
not in agreement with even the simplest infall model, but given the
observed brightness profiles of the protostars it is tempting to just
approximate the density distribution with a single power law. As an
example \cite{shirley00}, used this approach citing the results of
\cite{adams91}: if the density distribution \emph{can} be described by
a power-law and the beam can be approximated by a gaussian then the
outcoming brightness profiles will also be a power-law, so a power-law
fit to the outer parts of the brightness profile will directly reflect
the density distribution. This approach is, however, subject to noise in
the data and the parts of the brightness profile chosen to be
considered. On the other hand, our 1D modelling clearly shows that the
data do not warrant more complicated fits and that the power-law
adequately describes the profiles of the sources. Modelling of
the detailed line profiles will require more sophisticated infall
models, since signatures for infall exist for around half of the class
0 sources in the sample \citep{andreppiv}. However, \cite{schoier02}
show that for the case of IRAS 16293-2422, adopting the infall model
of \cite{shu77} does not improve the quality of the fit to the
continuum data.

\begin{figure}
\resizebox{\hsize}{!}{\rotatebox{90}{\includegraphics{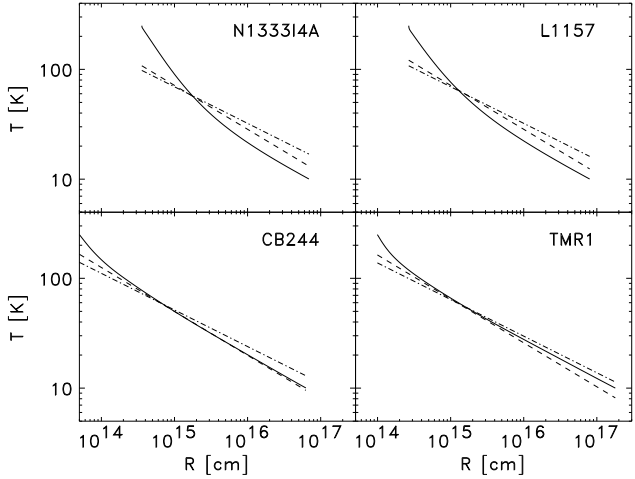}}}
\caption{The temperature profiles for four selected sources with
temperature profiles calculated in the Rayleigh-Jeans limit for the
optically thin assumption overplotted for different dust opacity laws,
$\kappa_\nu\propto \nu^\beta$. The dashed line corresponds to
$\beta=1$ and the dash-dotted line to $\beta=2$.}\label{tempprofiles}
\end{figure}

\cite{chandler00} assumed that the envelopes were optically thin. In
this case, the temperature profile can be shown to be a simple
power-law as well and \citeauthor{chandler00} subsequently derived
analytical models which could be fitted directly to the SCUBA
data. For the sources included in both samples the derived power-law
indices agree within the uncertainties. However, as illustrated in
Fig.~\ref{tempprofiles}, the optically thin assumption for the
temperature distribution is not valid in the inner parts of the
envelope, especially of the more massive class 0 sources, so actual
radiative transfer modelling is needed to establish the temperature
profile, crucial for calculations of the molecular excitation and
chemical modelling.

Disk emission can contribute to the fluxes of the innermost points on
the brightness profiles and thus lead to a steeper density
profile. This is likely to be more important in the sources with the
less massive envelope, i.e. class I objects. In tests where the fluxes
within the innermost 15\arcsec\ of the brightness profiles are reduced
by 50\%, the best fit values of $\alpha$ are reduced by 0.1--0.2. This
is comparable to the uncertainties in the derived value of $\alpha$,
but can introduce a systematic error.

It is interesting to note that there is no clear trend in the slope of
the density profile with type of object. In the framework of the
inside-out collapse model, one would expect a flattening of the
density profiles, approaching 1.5 as the entire envelope undergoes the
collapse. This is not seen in the data - actually the average density
profile for the class I objects is slightly steeper than for the class
0 objects. On the other hand, the suggestion of an outer envelope with
a flat density distribution and a significant fraction of material as
suggested by \cite{jayawardhana01} can also not be confirmed by this
modelling. As seen from the fits in Fig.~\ref{bright450sum} and
\ref{bright850sum} the brightness profiles only suggest departures
from the single power-law fits in the outer regions in a few cases, so
if such a component is present, it is not traced directly by the SCUBA
maps. The slightly flatter density profiles for the class 0 objects
could be a manifestation of such an outer component - but the density
distributions of the sources in this sample are typically much steeper
(i.e. $\rho \propto r^{-3/2}$) than those modelled by
\citeauthor{jayawardhana01} ($\rho \propto r^{-1/2}$).

\subsection{Geometrical effects}\label{geometry}
As mentioned above, it was assumed in the modelling that the sources
are spherical symmetric. This may not be entirely correct considering
the structure of YSOs, which may be rotating and are permeated by
magnetic fields leading to polar flattening \citep{terebey84,galli93}
and which are definitely associated with molecular outflows and jets
\citep{bachillercreteII,richerppiv}.  One problem in this discussion
is the exact shape of the error beam of SCUBA. Typically an error lobe
pickup of 15\% in the 850~$\mu$m images and 45\% in the 450~$\mu$m
maps is estimated. This error beam is not completely spherically
symmetric, but is also not well-established, so in fact 1D modelling
may be the best that can be done using the SCUBA data.

\cite{myers98} investigated the results of departure from spherical
symmetry of an envelope when calculating the bolometric temperature of
YSOs seen under various inclination angles. With a cavity in the pole
region, roughly corresponding to the effect of a bipolar outflow, they
found that the bolometric temperature could increase by a factor
1.3--2.5 for a typical opening angle of 25\degr. A similar line of
thought can be applied to our modelling: departure from spherical
symmetry by having a thinner polar region will affect the
determination of the SED - a source viewed more pole-on will see
warmer material which leads to an SED shifted towards shorter
wavelengths and accordingly a lower value of $\tau_{100}$.
\citeauthor{myers98} also argued that the effect on a statistical
sample would be rather small, e.g., compared to differences in optical
depth, but for studies of individual sources like in our case, this
effect might be of importance.

The brightness profile will also change in the aspherical case. In the
case of a source viewed edge-on this would result in elliptically
shaped SCUBA images with the 850~$\mu$m data showing a more elongated
structure since the smaller optical depths material in the polar
regions would reveal material being warmer and thus having stronger
emission at 450~$\mu$m, compensating for the lack of material in these
images. If we consider the case where the source is viewed entirely
pole-on, the image would still appear circular, but the brightness
profiles would show a steeper increase towards the center in the
450~$\mu$m data (closer to the spherical case) than the 850~$\mu$m
data for the same reasons as mentioned above.

The good fits to the brightness profiles given the often non-circular
nature of the SCUBA images is expected based on the mathematical
nature of power-law profiles. Consider as an example a 2D image of a
source described by:
\begin{equation}
I=I_0\, r^{-f(\theta)} 
\end{equation}
where $f(\theta)$ is a function describing the variation of the slope
of the brightness profiles extracted in rays along different
directions away from the center position. In the case of a source with
a simple density profile $\rho\propto r^{-p}$ \cite{adams91} showed
that such sources will also give images with power-law brightness
distributions corresponding to the case where $f(\theta)={\rm
const}$. As a somewhat simplistic case, assume that $f(\theta)$ is a
step function with a power-law slope $p_1$ in $\theta \in [0,\pi[$ and
a $p_2$ in $\theta \in [\pi,2\pi[$. In this case determination of the
power-law slope will be dominated by the flattest of the two slopes:
the azimuthal averaged brightness profile will be
\begin{equation}
\langle I\rangle_{\theta}\propto r^{-p_1}+r^{-p_2}
\end{equation}
and due to the power-law decline the term corresponding to the
shallower slope will dominate the average, especially at the larger
radii. This has two effects: first the distribution of power-law
``rays'' must be very strongly varying in order not to result in a
power-law average, and second, the density profile for asymmetric
sources will be flattened compared to more spherical sources.

As a test, brightness profiles were extracted in angles covering
respectively the flattest and steepest direction of \object{L483} and
\object{L1527}. Modelling the so-derived brightness profiles gives
best-fit density profiles of 0.9 and 1.2, compared to the 0.9 derived
as the average over the entire image for \object{L483}, and 0.6 and
0.8 for L1527 compared to 0.6 from the entire image. Thus, the
brightness profile and the derived density profile might indeed be
flattened by the asymmetry and could account for the somewhat flatter
profiles found towards some sources. The discrepancies between the
profiles along different directions are, however, not much larger than
the uncertainties in the derived power-law slope, so these sources
could have intrinsic flatter density distributions instead; with the
present quality of the data both interpretations are possible.

\subsection{Pre-stellar cores}\label{ppcs}
The modelling of the pre-stellar cores is more complicated than that
of the class 0 and I sources, since it is not clear whether the cores
are undergoing gravitational collapse, or are centrally condensed
and/or are gravitationally bound. In the case of thermally supported
gravitationally bound cores, the solution for the density profile is
the Bonnor-Ebert sphere \citep{ebert55,bonnor56}. Recently
\cite{evans01} modelled three pre-stellar cores (including
\object{L1689B} and \object{L1544} in our sample) and found that they
could be well fitted by Bonnor-Ebert spheres. \citeauthor{evans01}
also found that the denser cores, \object{L1689B} and \object{L1544},
were those showing spectroscopic signs of contraction thus suggesting
an evolutionary sequence with \object{L1544} as the pre-stellar cores
closest to the collapse phase. This is supported as well by millimeter
observations of this core which show a dense inner region
\citep{tafalla98,wardthompson99}.

\citeauthor{evans01} also discussed the properties of non-isothermal
models and found that the quality of the fit to the data is similar to
the isothermal case. \cite{wardthompson02} examined ISOPHOT 200 $\mu$m
data for a sample of pre-stellar cores (including \object{L1689B} and
\object{L1544}) and found that none of the cores had a central peak in
temperature and that they could all be interpreted as being isothermal
or having a temperature gradient with a cold center as result of
external heating by the interstellar radiation field. \cite{zucconi01}
derived analytical formulae for the dust temperature distributions in
pre-stellar cores showing that these cores should have temperatures
varying from 8~K in the center to around 15~K at the boundary. These
equations will be useful for more detailed modelling of the continuum
and line data, but for the present purpose the isothermal models are
sufficient.

Modelling the pre-stellar cores using our method is not possible as is
illustrated by the best fits for the pre-stellar cores shown in
Fig.~\ref{bright450sum}-\ref{sedsum}. Given the observational evidence
that these cores do not have central source of heating, what is
implicitly assumed in the DUSTY modelling, it is on the other hand
comforting that our method indeed distinguishes between these
pre-stellar cores and the class 0/I sources and that the obtained fits
to the brightness profiles of the latter sources are not just the
results of, e.g., the convolution of the ``real'' brightness profiles
with the SCUBA beam.

For modelling of sources without central heating \citeauthor{evans01}
note that their modelling does not rule out a power-law envelope
density distribution for \object{L1544}, as opposed to, e.g., the case
for \object{L1689B}: this would imply an evolutionary trend of the
pre-stellar cores having a Bonnor-Ebert density distribution, which
would then evolve towards a power-law density distribution with an
increasing slope as the collapse progresses. In the modelling of
spectral line data for the pre-stellar cores we adopt an isothermal
Bonnor-Ebert sphere with $n_c \sim 10^6$ for both \object{L1689B} and
\object{L1544} as this was the best fitted isothermal model in the
work of \cite{evans01}.

\section{Monte Carlo modelling of CO lines}\label{gasmodelling}
\subsection{Method}
One main goal of our work is to use the derived physical models as
input for modelling the chemical abundances of the various molecules
in the envelopes. To demonstrate this approach, modelling of the first
few molecules, C$^{18}$O and C$^{17}$O, is presented here. This
modelling also serves as a test of the trustworthiness of the physical
models: is it possible to reproduce realistic abundances for the
modelled molecules?

The 1D Monte Carlo code developed by \cite{hogerheijde00vandertak} is
used together with the revised collisional rate coefficients from
\cite{flower01}, and a constant fractional abundance over the entire
range of the envelope is assumed as a first approximation. Furthermore
the dust and gas temperatures are assumed identical over the entire
envelope and any systematic velocity field is neglected. In the outer
parts of the envelopes, the coupling between gas and dust may break
down \citep[e.g.][]{ceccarelli96,doty97} leading to differences
between gas and dust temperatures of up to a factor of 2.

\begin{figure}
\resizebox{\hsize}{!}{\includegraphics{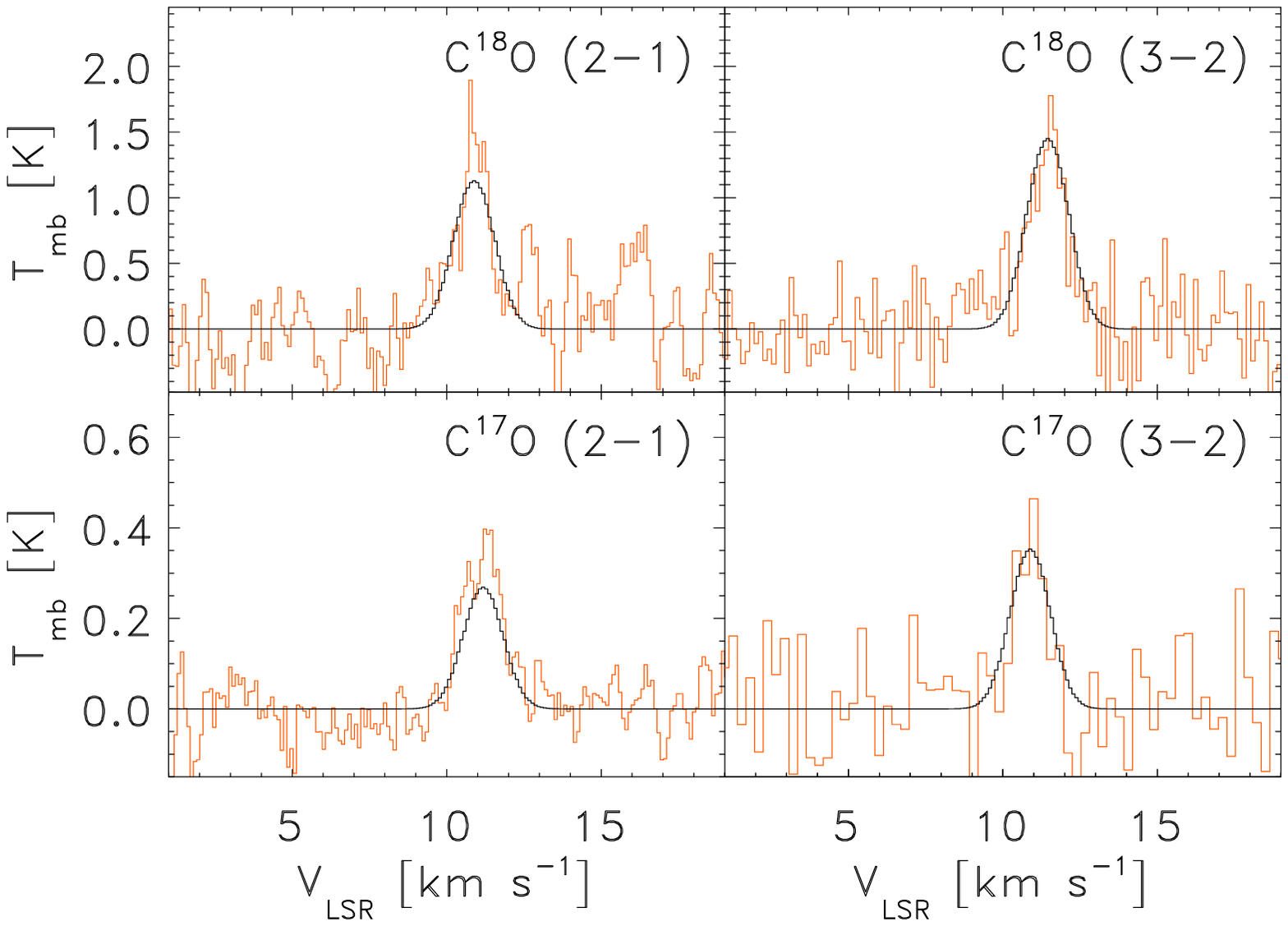}}
\resizebox{\hsize}{!}{\includegraphics{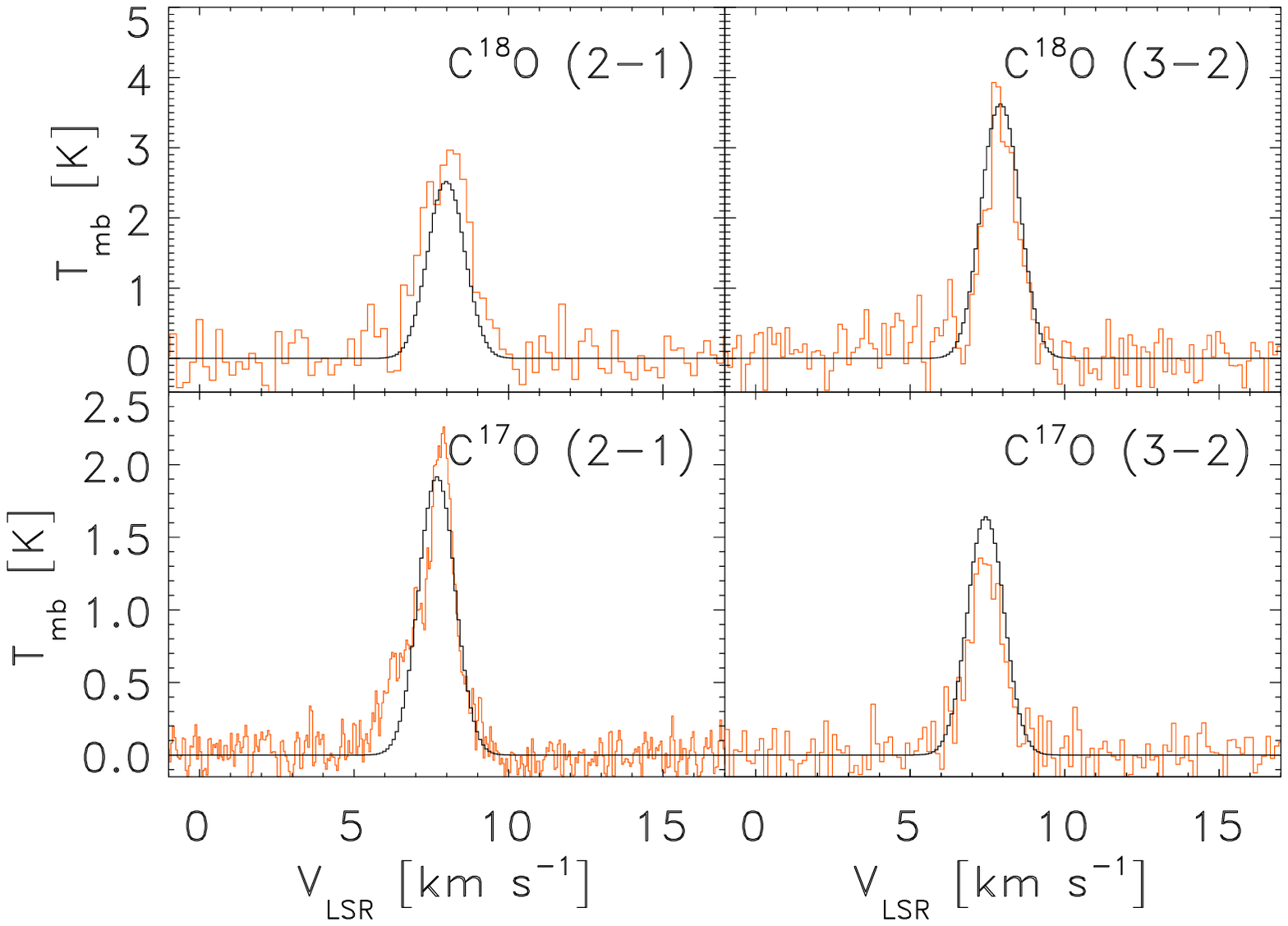}}
\caption{The C$^{18}$O and C$^{17}$O spectra for \object{L723} (upper
four panels) and \object{N1333-I2} (lower four panels) overplotted
with the best fit models from the Monte Carlo
modelling.}\label{modexample}
\end{figure}

With the given physical model and assumed molecular properties, there
are two free parameters which can be adjusted by minimizing $\chi^2$
to model the line profiles for each molecular transition: the
fractional abundance [X/H$_2$] and the turbulent line width $V_{\rm
D}$. Since emission from the ambient molecular cloud might contribute
to the lower lying 1--0 transitions, the derived abundances are based
only on fits to the 2--1 and 3--2 lines. The observed and modelled
line intensities are summarized in
Tables~\ref{line_int_jcmt}-\ref{line_int_cso}, whereas the parameters
for the best fit models are given in Table~\ref{abund_fits}. While a
turbulent line width of $0.5-1.0$ km~s$^{-1}$ is needed for most
sources to fit the actual line profile, the modelled line strengths
are only weakly dependent on this parameter compared to the fractional
abundance. For example, for \object{L723} one derives C$^{18}$O
abundances between $3.8 \times 10^{-8}$ and $4.0 \times 10^{-8}$ from
respectively $V_{\rm D}=1.1$ km~s$^{-1}$ (FWHM $\simeq 1.9$ km
s$^{-1}$) to $V_{\rm D}=0.7$ (FWHM $\simeq 1.2$ km~s$^{-1}$), which
should be compared to the [C$^{18}$O/H$_2$]$=3.9 \times 10^{-8}$ and
$V_{\rm D}=0.8$ km~s$^{-1}$ (FWHM $\simeq 1.4$ km~s$^{-1}$) quoted in
Table~\ref{abund_fits}. The C$^{18}$O and C$^{17}$O spectra for
\object{L723} and \object{N1333-I2} with the best fit models
overplotted are shown in Fig.~\ref{modexample}. The revised
collisional rate coefficients from \cite{flower01} adopted for this
modelling typically change the derived abundances by less than 5\%
compared to results obtained from simulations with previously
published molecular data. The relative intensities of the various
transitions and so the quality of the fits remain the same.

As it is evident from Tables~\ref{line_int_jcmt}-\ref{line_int_cso},
the 2--1 and 3--2 lines can be well modelled using the above approach,
while the 1--0 lines are significantly underestimated from the
modelling, especially in the larger Onsala 20m beam. This indicates
that the molecular cloud material may contribute significantly to the
observations or that the assumed outer radius is too small. The
importance of the latter effect was tested by increasing the outer
radius of up to a factor 2.5 for a few sources and it was found that
while the 1--0 and 2--1 line intensities could increase in some cases
by up to a factor of 2, the 3--2 line intensities varied by less than
10\%, illustrating that the 3--2 line mainly trace the warmer ($\ge$
30~K) envelope material.

The importance of the size of the inner radius was tested as
well. For \object{L723} fixing the inner radius at 50~AU rather
than 8~AU increases the best fit abundance by $\sim$ 5\% to
[C$^{18}$O/H$_2$]$=4.1 \times 10^{-8}$ without changing the quality of
the fit significantly (drop in $\chi^2$ of $\sim$ 0.1).  This is also
found for other sources in the sample and simply illustrates that
increasing the inner radius corresponds to a (small) decrease in mass
of the envelope, so that a higher CO abundance is required to give the
same CO intensities.  Yet, it is good to keep in mind that the inner
radius of the envelope may be different, and even though its value
does not change the results for CO, it might for other molecules,
e.g., CH$_3$OH and H$_2$CO, which trace the inner warm and dense
region of the envelope.
\begin{table*}
\caption{The observed C$^{18}$O and C$^{17}$O lines from the JCMT
observations compared to the modelled
intensities.}\label{line_int_jcmt}
\begin{center}
\begin{tabular}{lllllll}\hline\hline
Source       & $\Delta V_{\rm av}$ & $V_{\rm LSR}$ & \multicolumn{2}{c}{(2--1)} &\multicolumn{2}{c}{(3--2)} \\ 
             & (km~s$^{-1}$)\fno{a} & (km~s$^{-1}$) & Obs\fno{b} & Mod\fno{c}  & Obs\fno{b} & Mod\fno{c} \\ \hline
\multicolumn{7}{c}{\rule[0.5ex]{0mm}{2.0ex}C$^{18}$O} \\ \hline
\object{L1448-I2}     & 0.7 & 4.5 & $\ldots$  & 1.6 & 1.9$\ast$ & 1.9 \\
\object{L1448-C}      & 1.2 & 5.3 & {4.3}     & 3.3 & 3.3$\ast$ & 3.9 \\
\object{N1333-I2}     & 1.4 & 7.7 & {5.8}     & 3.9 & 4.7$\ast$ & 5.7 \\
\object{N1333-I4A}    & 1.7 & 7.2 & {5.5}     & 5.4 & $\ldots$  & 7.1 \\
\object{N1333-I4B}    & 2.1 & 7.4 & {4.3}     & 3.8 & 4.0$\ast$ & 4.4 \\
\object{L1527}        & 0.7 & 5.9 & {4.4}     & 2.9 & {2.3}     & 2.7 \\
\object{VLA1623}      & 1.0 & 3.5 & {12.1:}   & 12.1 & {12.6:}  & 12.6 \\ 
\object{L483}         & 1.0 & 5.2 & {4.1}     & 3.7 & {3.6}     & 3.8 \\ 
\object{L723}         & 1.6 & 11.2 & {2.2}    & 1.9 & 2.2$\ast$ & 2.4 \\ 
\object{L1157}        & 0.8 & 2.6 & {1.7}     & 1.2 & {1.4}     & 1.7 \\ 
\object{CB244}        & 1.1 & 4.4 & {3.3}     & 2.9 & {2.9}     & 3.1 \\
\object{L1489}        & 2.1 & 7.2 & {2.7}     & 2.6 & {3.8}     & 3.9 \\
\object{TMR1}         & 1.5 & 6.3 & {4.0}     & 3.3 & {3.8}     & 4.3 \\
\object{L1551-I5}     & 2.1 & 7.2 & {7.1}     & 6.1 & {8.7}     & 9.7 \\
\object{TMC1A}        & 1.5 & 6.6 & {1.3}     & 1.5 & {2.8}     & 2.3 \\
\object{TMC1}         & 1.5 & 5.2 & {2.3}     & 2.7 & {4.3}     & 3.2 \\
\object{L1544}        & 0.3 & 7.6 & $\ldots$  & 0.55 & {0.30}     & 0.30 \\
\object{L1689B}       & 0.6 & 3.6 & {3.6}     & 3.3 & {1.8}     & 1.9 \\ \hline
\multicolumn{7}{c}{\rule[0.5ex]{0mm}{2.0ex}C$^{17}$O} \\ \hline
\object{L1448-I2}     & 0.9 & 4.0 & $\ldots$       & 0.61 & 0.71$\ast$ & 0.72 \\
\object{L1448-C}      & 1.4 & 5.0 & {1.5}          & 1.4  & 1.7$\ast$ & 1.7 \\
\object{N1333-I2}     & 1.3 & 7.5 & $\ldots$       & 1.4  & 1.8$\ast$ & 2.1 \\
\object{N1333-I4A}    & 1.3 & 6.7 & $\ldots$       & 1.1  & 1.4$\ast$ & 1.6 \\
\object{N1333-I4B}    & 1.4 & 6.8 & $\ldots$       & 0.9  & $\ldots$  & 1.1 \\
\object{L1527}        & 0.5 & 6.0 & {1.9}          & 1.7  & {1.3}     & 1.6 \\
\object{VLA1623}      & 0.8 & 3.7 & $\ldots$       & 3.8  & 4.6:$\ast$& 4.5 \\
\object{L483}         & 0.8 & 5.3 & {1.5}          & 1.3  & 1.2$\ast$ & 1.3 \\
\object{L723}         & 1.3 & 11.0 & {0.59}        & 0.43 & 0.49$\ast$ & 0.57 \\
\object{L1157}        & 1.0 & 2.7 & {0.51}         & 0.47 & 0.65$\ast$ & 0.69 \\
\object{CB244}        & 1.0 & 4.2 & {0.94}         & 0.68 & 0.60$\ast$ & 0.73 \\
\object{L1489}        & 3.0 & 7.3 & $\ldots$       & 0.56 & 0.86$\ast$ & 0.88 \\
\object{TMR1}         & 1.6 & 6.0 & $\ldots$       & 0.69 & 0.84$\ast$ & 0.94 \\
\object{L1551-I5}     & 2.1 & 7.2 & {2.2}          & 1.8  & {2.6}     & 3.1 \\
\object{TMC1A}        & 1.5 & 6.6 & $\ldots$       & 0.45 & $\ldots$   & 0.73 \\
\object{TMC1}         & 1.5 & 5.2 & $\ldots$       & 0.83 & $\ldots$   & 1.0 \\
\object{L1544}        & 0.6 & 7.5 & {0.26}         & 0.26 & $\ast$\fno{d}   & 0.14 \\
\object{L1689B}       & 0.6 & 3.7 & {0.71}         & 0.85 & 0.59$\ast$ & 0.44 \\ \hline
\end{tabular}
\end{center}
Notes: \fno{a}The width (FWHM) of the lines. \fno{b}Observed
intensities, $\int T_{\rm mb}$ d$V$ in K km s$^{-1}$: our own
observations are marked with '$\ast$' and lines where double gaussians
were fitted and the broadest component subtracted marked with
':'. \fno{c}Modelled intensities in K km~s$^{-1}$. \fno{d}C$^{17}$O
3--2 not detected towards \object{L1544} in integrations corresponding
to an RMS of 0.1~K ($T_{A}^{\ast}$).
\end{table*}
\begin{table}
\caption{As in Table~\ref{line_int_jcmt} but for the IRAM 30m
observations - all from November 2001 observing
run.}\label{line_int_iram}
\begin{center}
\begin{tabular}{lllll}\hline\hline
Source    & \multicolumn{2}{c}{C$^{17}$O 1--0} & \multicolumn{2}{c}{C$^{17}$O 2--1} \\
          & Obs & Mod  & Obs  & Mod \\ \hline
\object{L1448-C}   & 0.58 & 0.71 & 2.2 & 2.3 \\
\object{N1333-I2}  & 1.1  & 0.61 & 3.4 & 2.5 \\
\object{N1333-I4A} & 0.99 & 0.54 & 2.9 & 2.3 \\
\object{N1333-I4B} & 0.73 & 0.47 & 1.8 & 1.5 \\
\object{L1527}     & 1.1  & 0.90 & 2.4 & 2.0 \\
\object{L1489}     & 0.37 & 0.27 & 1.1 & 1.1 \\
\object{TMR1}      & 0.69 & 0.37 & 1.4 & 1.3 \\ \hline
\end{tabular}
\end{center}
Notes: The intensities are the total observed intensity summed over
the hyperfine-structure lines.
\end{table}
\begin{table}
\caption{As in Table~\ref{line_int_jcmt} and \ref{line_int_iram} for
the Onsala 20m observations - all from March 2002 observing
run.}\label{line_int_onsala}
\begin{center}
\begin{tabular}{lllll}\hline\hline
Source    & \multicolumn{2}{c}{C$^{18}$O 1--0} & \multicolumn{2}{c}{C$^{17}$O 1--0} \\
                   & Obs  & Mod  & Obs  & Mod \\ \hline
\object{L1448-I2}  & 3.2  & 0.5 & 1.1 & 0.2 \\
\object{L1448-C}   & 3.6  & 1.3  & $\ldots$ & 0.5 \\
\object{N1333-I2}  & 5.6  & 1.4  & 1.6 & 0.5 \\
\object{N1333-I4A} & 4.2  & 1.9  & $\ldots$ & 0.4 \\
\object{N1333-I4B} & 4.1  & 1.5  & $\ldots$ & 0.4 \\
\object{L1527}     & 2.4  & 1.5  & $\ldots$ & 0.8 \\
\object{L483}      & 3.8  & 1.7  & 2.2 & 0.6 \\
\object{L723}      & 1.6  & 0.6  & 0.3 & 0.1 \\
\object{L1157}     & 1.3  & 0.3  & 0.2 & 0.1 \\
\object{CB244}     & 1.9  & 1.1  & $\ldots$ & 0.2 \\
\object{L1489}     & 1.7  & 1.0  & $\ldots$ & 0.2 \\
\object{TMR1}      & 1.7  & 1.5  & $\ldots$ & 0.3 \\
\object{L1544}     & 1.7  & 0.3 & 0.6 & 0.2 \\ \hline
\end{tabular}
\end{center}
Notes: \fno{a}The intensities are the total observed intensity summed over
the hyperfine-structure lines.
\end{table}
\begin{table}
\caption{As Tables~\ref{line_int_jcmt}--\ref{line_int_onsala} for
CSO measurements from the literature.}\label{line_int_cso}
\begin{center}
\begin{tabular}{lllll}\hline\hline
Source     & \multicolumn{2}{c}{C$^{18}$O 3--2\fno{a}} & \multicolumn{2}{c}{C$^{17}$O 3--2\fno{a}} \\
           & Obs  & Mod  & Obs  & Mod  \\ \hline
\object{N1333-I4A}  & 4.9 & 4.9 & 1.0  & 1.0  \\
\object{N1333-I4B}  & 3.3 & 3.3 & 0.80 & 0.80 \\ \hline
           & \multicolumn{2}{c}{C$^{17}$O 2--1\fno{b}} & & \\ \hline
\object{TMC1A}      & 0.28 & 0.27 & & \\
\object{TMC1}       & 0.53 & 0.51 & & \\ \hline
\end{tabular}
\end{center}
Notes: \fno{a}From \cite{blake95}, half-power beam width of
20\arcsec. \fno{b}From \cite{ladd98}, half-power beam width of
33\arcsec.
\end{table}

\subsection{CO abundances}
\begin{figure*}
\centering
\resizebox{\hsize}{!}{\includegraphics{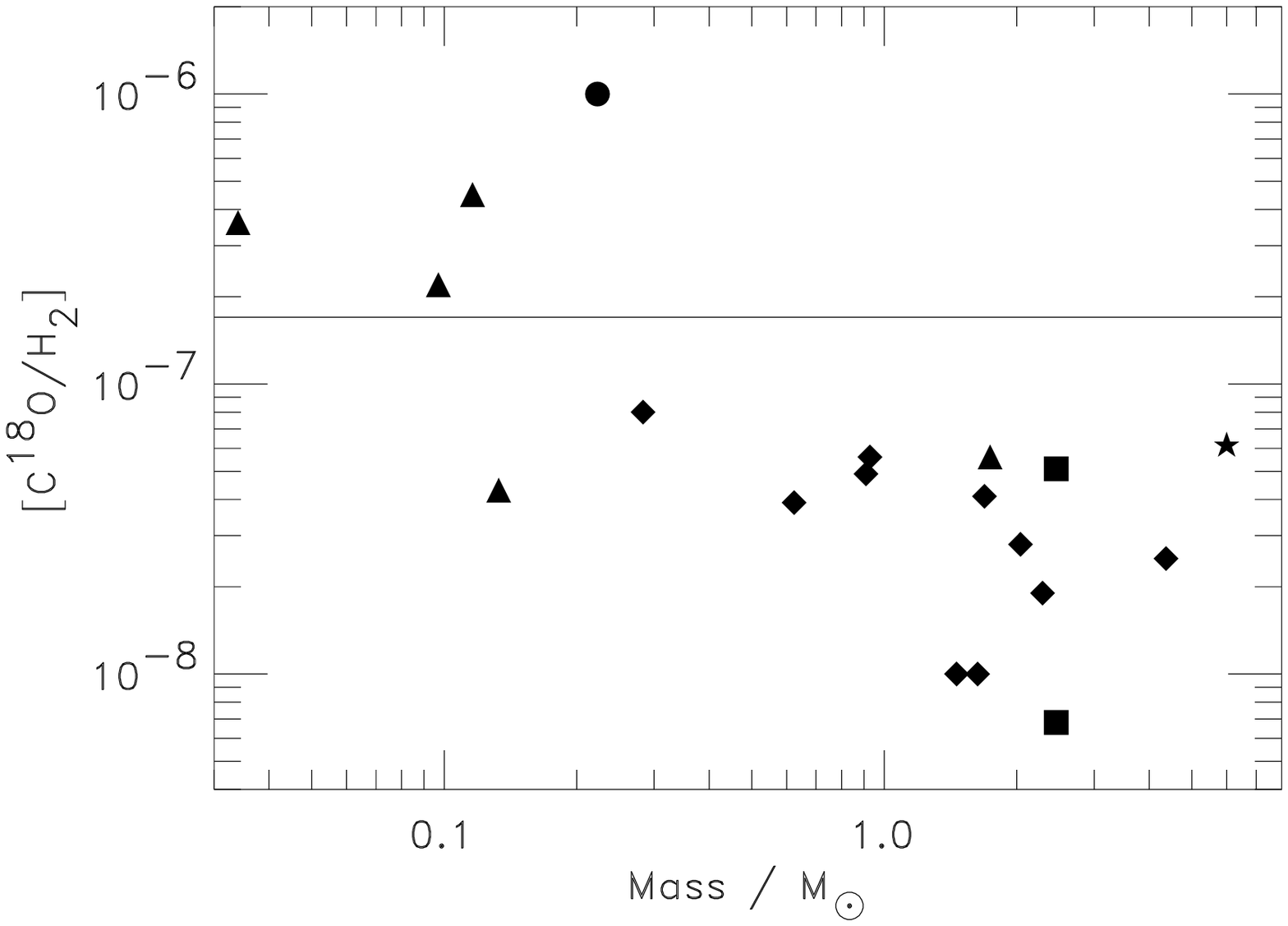}\includegraphics{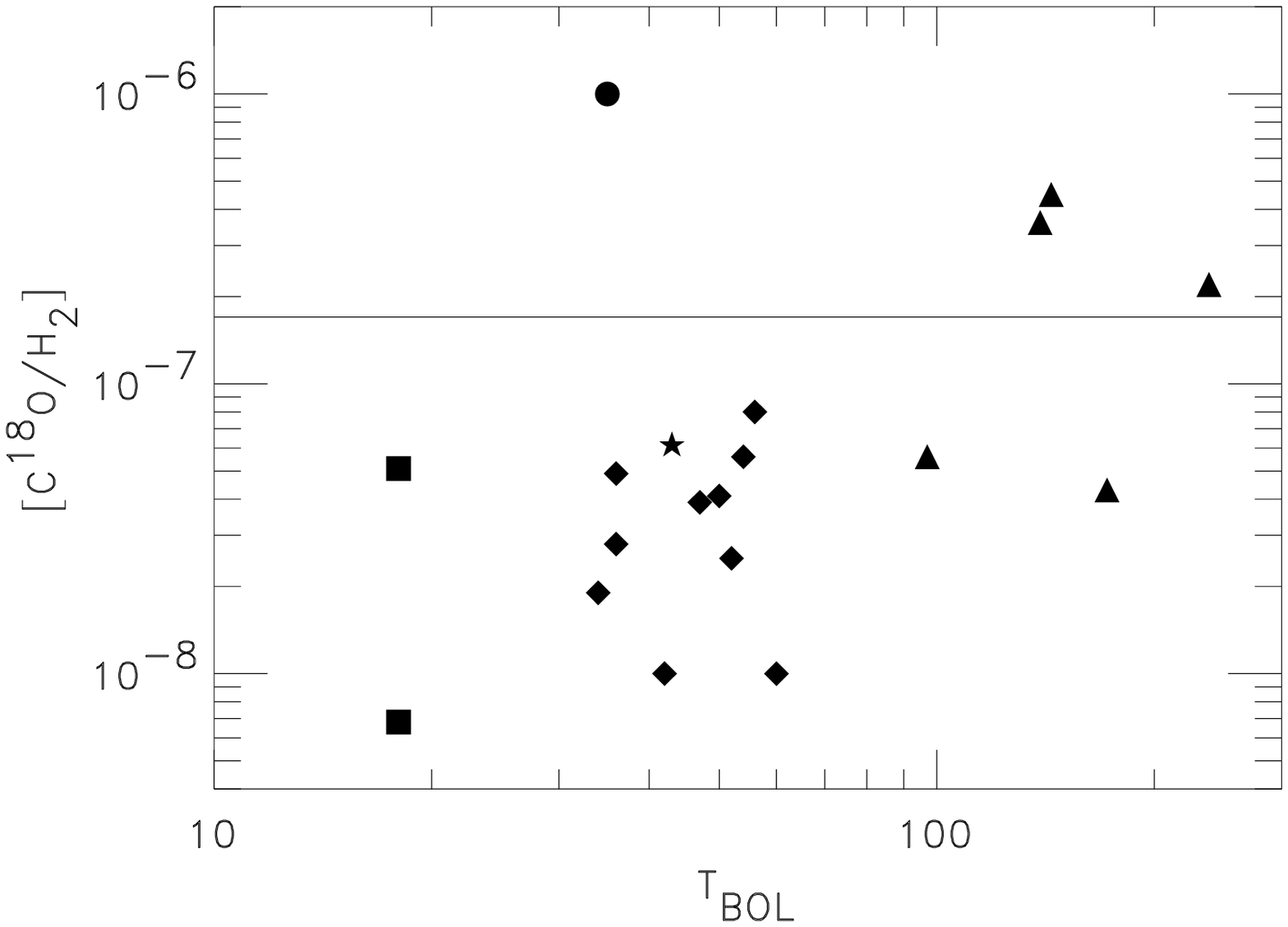}}
\resizebox{\hsize}{!}{\includegraphics{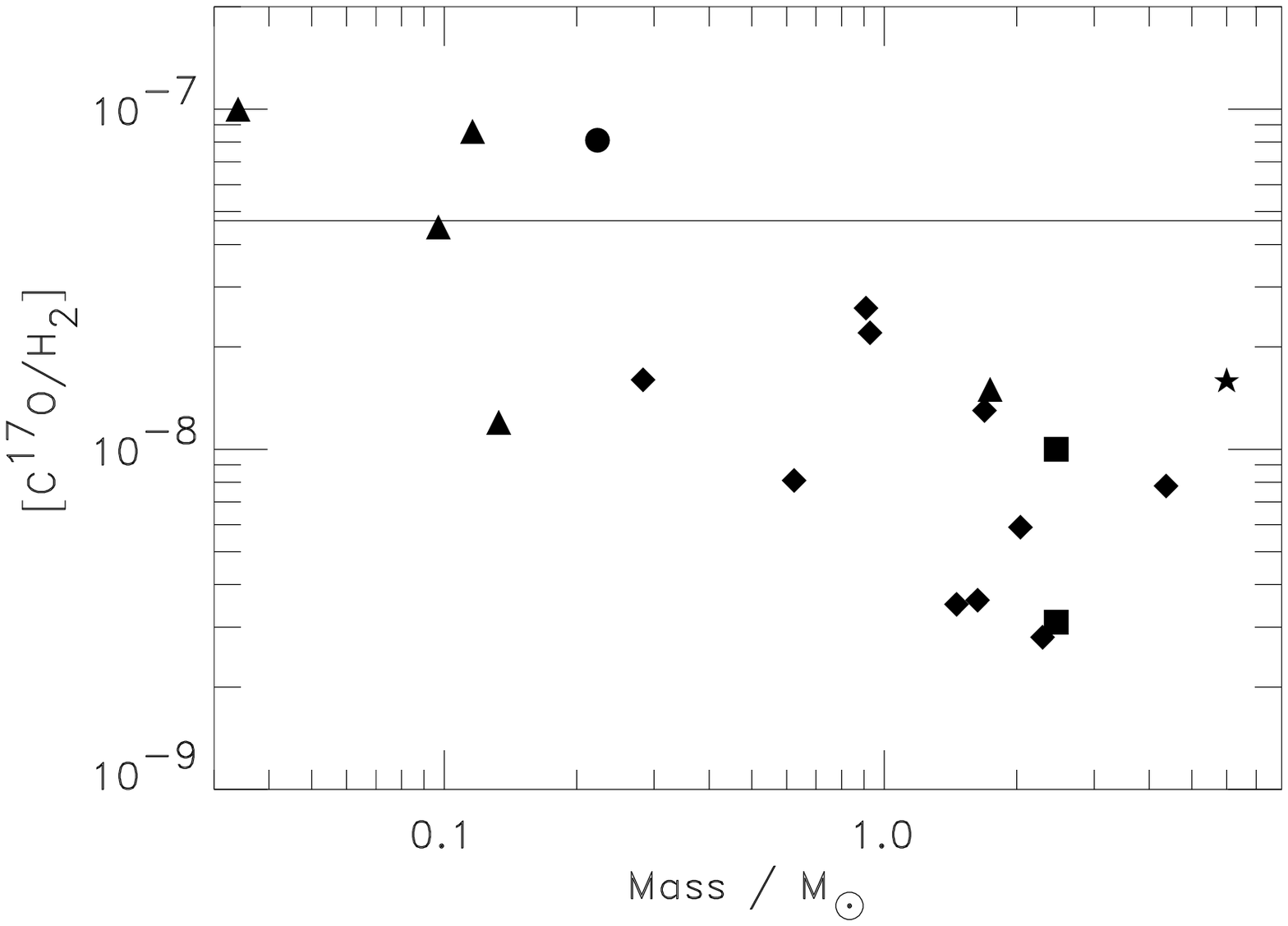}\includegraphics{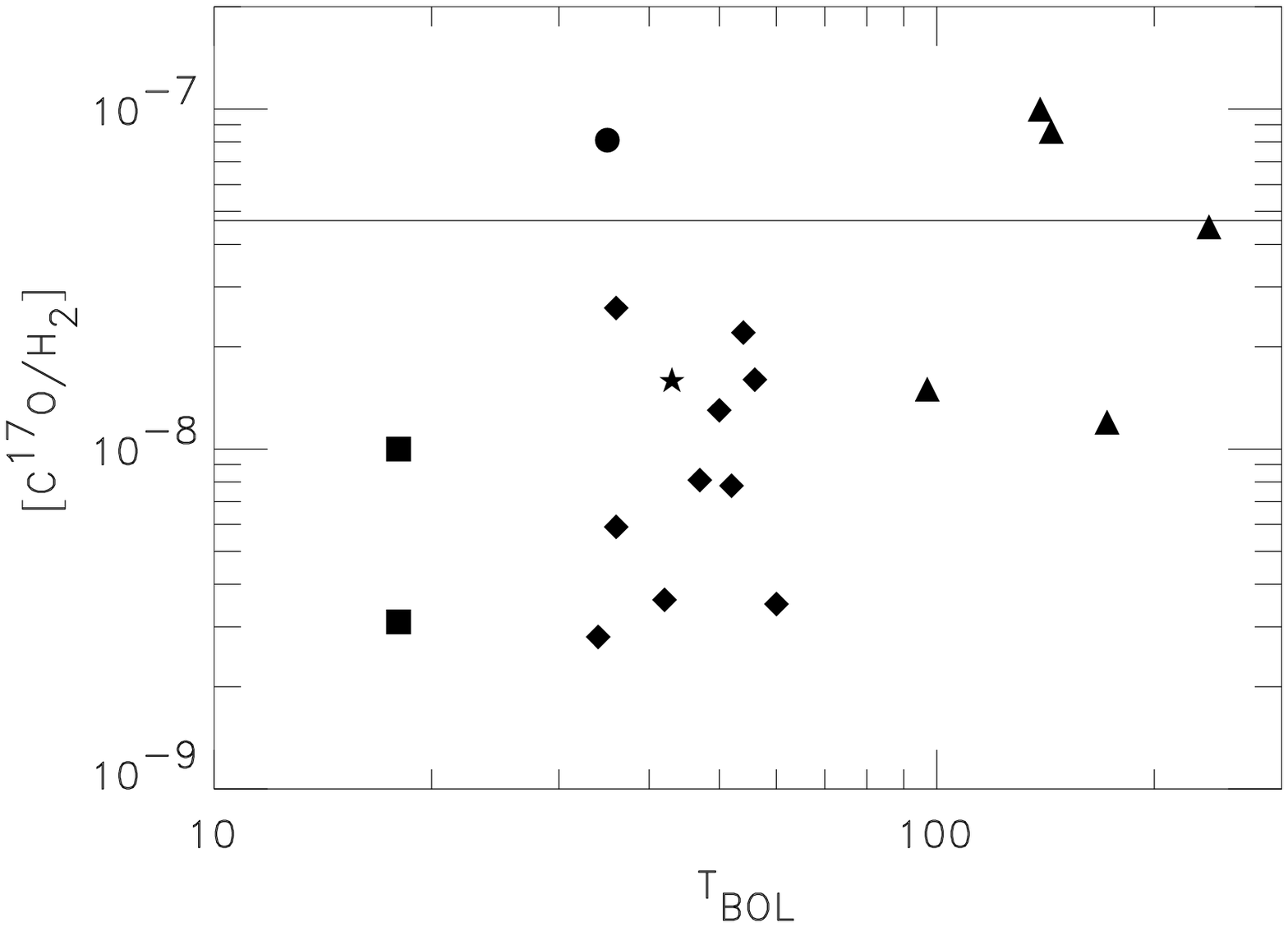}}
\caption{The fitted abundances vs. envelope mass (left) and bolometric
temperature (right) of each source for respectively the C$^{18}$O data
(\emph{top}) and C$^{17}$O data (\emph{bottom}). The sources have
been split into groups (class 0, class I and pre-stellar cores) with
VLA 1623 and IRAS 16293-2422 separated out using the same symbols as
in Fig.~\ref{tbolfysparam}: class 0 objects are marked with
``$\blacklozenge$'', class I objects with ``$\blacktriangle$'',
pre-stellar cores with ``$\blacksquare$'', VLA 1623 with ``{\large
$\bullet$}'' and IRAS 16293-2422 with ``$\bigstar$''. The vertical
lines in the figures illustrate the abundances in quiescent dark
clouds from \cite{frerking82}.}\label{massabund}
\end{figure*}

The fitted abundances are summarized in Table~\ref{abund_fits} and
plotted against the envelope mass of each individual source in
Fig.~\ref{massabund}. As can be seen from the values of the reduced
$\chi^2$ for the fit to the data for each isotope, the model
reproduces the excitation of the individual species. Even in the
``worst case'' of \object{N1333-I2}, the lines do indeed seem to be
well-fitted by the model (see Fig.~\ref{modexample}). One source,
\object{VLA1623} shows a remarkably high ratio between the C$^{18}$O
and C$^{17}$O abundances of 12.4 and high C$^{18}$O and C$^{17}$O
abundances in comparison to the rest of the class 0 sources. Given the
location of VLA 1623 close to a dense ridge of material and the
associated uncertainties in the physical models of this source, it may
reflect problems in the model rather than being a real property of the
source. For the rest of the sources, the ratio between the C$^{18}$O
and C$^{17}$O abundances is found to be $3.9\pm 1.3$ in agreement with
the expected value from, e.g., the local interstellar medium of 3.65
\citep{penzias81,wilson94} - another sign that the model reproduces
the physical structure of the envelopes and that no systematic
calibration errors are introduced by using data from the various
telescopes and receivers.

It is evident that the class 0 objects (except VLA 1623) and
pre-stellar cores show a high degree of depletion compared to the
expected abundances of [C$^{18}$O/H$_2$] of $1.7\times 10^{-7}$ from
nearby dark clouds \citep{frerking82} and [C$^{17}$O/H$_2$] of
$4.7\times 10^{-8}$ assuming $^{18}$O:$^{17}$O of 3.65. With this
isotope ratio and assuming $^{16}$O:$^{18}$O equal to 540
\citep{wilson94}, the average abundance for the class I sources is
$(1.1\pm 0.9)\times 10^{-4}$ and for the class 0 sources and
pre-stellar cores $(2.0\pm 1.3)\times 10^{-5}$. The error bars
illustrate the source to source variations and uncertainties in
classifying borderline class 0/I objects like \object{CB244},
\object{L1527} and \object{L1551-I5}. Previously \cite{caselli99}
derived the C$^{17}$O abundance for one of the pre-stellar cores,
\object{L1544}, and our abundance agrees with their estimate within
the uncertainties. It is interesting to see that the class I objects
have higher CO abundances close to the molecular cloud values,
indicating that the class 0 objects indeed seem to be closer related
to the pre-stellar cores in this sense. Van der Tak et
al.~\citeyearpar{vandertak00} likewise found a trend of increasing CO
abundance with mass-weighted temperature for a sample of high-mass
YSOs and suggested that this trend was due to freeze-out of CO in the
cold objects.

\begin{table*}
\caption{CO abundances using the best models of Table~\ref{dustyphyspar}.}\label{abund_fits} 
\begin{center}
\begin{tabular}{llllllll}\hline\hline
Source & $v_{\rm D}$\fno{a} & [C$^{18}$O/H$_2$] & $\chi_{\rm red}^2$ & [C$^{17}$O/H$_2$] & $\chi_{\rm red}^2$ & $^{18}$O/$^{17}$O & [CO/H$_2$]\fno{c} \\ \hline
\object{L1448-I2}   & 0.5 & 1.0$\times 10^{-8}$ & $\ldots$ & 3.5$\times 10^{-9}$ & $\ldots$ & 2.9  & 6.1$\times 10^{-6}$ \\
\object{L1448-C}    & 0.7 & 5.6$\times 10^{-8}$ & 2.1      & 2.2$\times 10^{-8}$ & 0.07     & 2.5  & 3.7$\times 10^{-5}$ \\
\object{N1333-I2}   & 0.8 & 4.1$\times 10^{-8}$ & 3.9      & 1.3$\times 10^{-8}$ & 2.5      & 3.2  & 2.4$\times 10^{-5}$ \\
\object{N1333-I4A}  & 0.7 & 1.9$\times 10^{-8}$ & $<0.01$  & 2.8$\times 10^{-9}$ & 0.7      & 6.8  & 7.9$\times 10^{-6}$ \\
\object{N1333-I4B}  & 0.6 & 2.8$\times 10^{-8}$ & 0.30     & 5.9$\times 10^{-9}$ & $<0.01$  & 4.7  & 1.3$\times 10^{-5}$ \\
\object{L1527}      & 0.4 & 4.9$\times 10^{-8}$ & 3.5      & 2.6$\times 10^{-8}$ & 1.0      & 1.8  & 3.9$\times 10^{-5}$ \\
\object{VLA1623}    & 0.6 & 1.0$\times 10^{-6}$ & $<0.01$  & 8.1$\times 10^{-8}$ & $\ldots$ & 12.3 & 1.6$\times 10^{-4}$ \fno{e} \\
\object{L483}       & 0.6 & 2.5$\times 10^{-8}$ & 0.28     & 7.8$\times 10^{-9}$ & 1.1      & 3.2  & 1.4$\times 10^{-5}$ \\
\object{L723}       & 0.8 & 3.9$\times 10^{-8}$ & 0.83     & 8.1$\times 10^{-9}$ & 2.5      & 4.8  & 1.9$\times 10^{-5}$ \\
\object{L1157}      & 0.6 & 1.0$\times 10^{-8}$ & 3.3      & 3.6$\times 10^{-9}$ & 0.23     & 2.8  & 6.2$\times 10^{-6}$ \\
\object{CB244}      & 0.6 & 8.0$\times 10^{-8}$ & 0.47     & 1.6$\times 10^{-8}$ & 3.1      & 5.0  & 3.7$\times 10^{-5}$ \\
\object{L1489}      & 1.2 & 2.2$\times 10^{-7}$ & 0.13     & 4.5$\times 10^{-8}$ & 0.04     & 4.6  & 1.0$\times 10^{-4}$ \\
\object{TMR1}       & 0.9 & 4.5$\times 10^{-7}$ & 1.4      & 8.6$\times 10^{-8}$ & 1.0      & 4.5  & 2.0$\times 10^{-4}$ \\
\object{L1551-I5}   & 0.9 & 5.6$\times 10^{-8}$ & 0.80     & 1.5$\times 10^{-8}$ & 1.3      & 3.7  & 3.0$\times 10^{-5}$ \\
\object{TMC1A}      & 0.7 & 4.3$\times 10^{-8}$ & 1.3      & 1.2$\times 10^{-8}$ & $\ldots$ & 3.6  & 2.3$\times 10^{-5}$ \\
\object{TMC1}       & 0.7 & 3.6$\times 10^{-7}$ & 2.1      & 1.0$\times 10^{-7}$ & $\ldots$ & 3.6  & 2.0$\times 10^{-4}$ \\
\object{L1544}      & 0.3 & 6.8$\times 10^{-9}$ & $\ldots$ & 3.1$\times 10^{-9}$ & $\ldots$ & 2.2  & 4.9$\times 10^{-6}$ \\
\object{L1689B}     & 0.5 & 5.1$\times 10^{-8}$ & 0.36     & 1.0$\times 10^{-8}$ & 2.45     & 5.1  & 2.4$\times 10^{-5}$ \\[1.5ex]
\object{IRAS16293-2422}\fno{f} & & 6.2$\times 10^{-8}$ &  & 1.6$\times 10^{-8}$ &     & 3.9 & 3.3$\times 10^{-5}$ \\ \hline
\end{tabular}
\end{center}
Notes: ``$\ldots$'' indicate abundances where only one line where
available to constrain the fit. \fno{a}Turbulent velocity in km
s$^{-1}$, \fno{b}The $^{18}$O/$^{17}$O isotope ratio (or
[C$^{18}$O/H$_2$]/[C$^{17}$O/H$_2$]) \fno{c}Derived CO main isotope
abundance averaged from the C$^{18}$O and C$^{17}$O measurements
assuming $^{18}$O/$^{17}$O of 3.65 and $^{16}$O/$^{18}$O of 540
\citep{penzias81,wilson94} \fno{e}Based on the C$^{17}$O measurements
only. \fno{f}IRAS 16293-2422 included for comparison; for details see
\cite{schoier02}.
\end{table*}

The derived abundances are uncertain due to several factors. First,
the physical model and its simplicity and flaws as discussed
above. Second, the assumed dust opacities will affect the results: the
dust opacities may change with varying density and temperature in the
envelope, and will depend on the amount of coagulation and formation
of ice mantles. These effects tend to increase $\kappa_\nu$ with
increasing densities or lower temperatures, which will lower the
derived mass and thus increase the abundances necessary to reproduce
the same line intensities. Comparing the models of dust opacities in
environments of different densities and types of ice mantles as given
by \cite{ossenkopf94} indicates, however, that such variations should
be less than a factor of two and so cannot explain the differences in
the derived CO abundances.

The unknown contribution to the lower lines from e.g. the ambient
molecular cloud may affect the interpretation. Even for the 2--1
transitions the cloud material may contribute, leading to higher line
intensities than predicted from the modelling. Indeed as judged from
Tables~\ref{line_int_jcmt}-\ref{line_int_cso} there is a trend that
while the intensities of 2--1 lines are underestimated by the
modelling, the 3--2 lines are overestimated. This then means that the
derived CO abundances are upper limits - at least for the warmer
regions traced by the 3--2 lines.

As noted above the gas temperature could be lower by up to a factor of
two in the outer parts of the envelope \citep{ceccarelli96,doty97},
but this again would mainly affect the lines tracing the outer cold
regions i.e. the 1--0 and 2--1 lines compared to the 3--2 lines. At
the same time the gas temperatures from these detailed models may not
be appropriate for our sample. Our sources all have lower luminosities
than modelled e.g. by \cite{ceccarelli96}, leading to envelopes that
are colder on average, so that the relative effects of the decoupling
of the gas and dust are smaller. Another effect is external heating,
which as discussed in Sect.~\ref{ppcs} may lead to an increase in the
dust and gas temperatures in the outer parts.

A point of concern is also whether steeper density gradients could
 change the inferred abundances in light of the discussion about
 geometrical effects (Sect.~\ref{geometry}). Steepening of the density
 distribution would tend to shift material closer to the center and so
 towards higher temperatures. This would correspondingly change the
 ratio between the 2--1 and 3--2 lines towards lower values (stronger
 3--2 lines) and more so for the C$^{18}$O data than the C$^{17}$O
 data, since it traces the outer less dense parts of the
 envelope. Altogether none of the effects considered change the
 conclusion that the abundances in the class 0 objects are lower than
 those found for the class I objects.

\begin{figure}
\resizebox{\hsize}{!}{\includegraphics{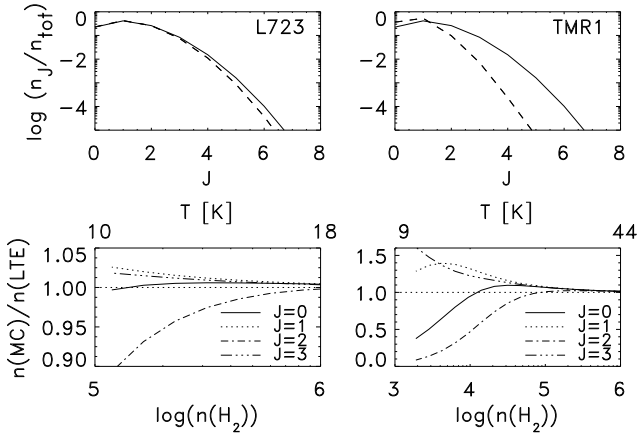}}
\caption{The derived level populations for C$^{18}$O in the envelopes
around the class 0 object \object{L723} (left column) and class I
object \object{TMR1} (right column). In the upper panels the level
populations from the modelling in the outer shell of the envelope are
shown (dashed line) together with the predictions from the LTE
assumptions (solid line). In the lower panels, the ratio of the
resulting level populations from the modelling and the LTE predictions
are shown with varying density for the $J=0$ to 3 levels.}\label{lte}
\end{figure}

One important conclusion of the derived abundances is that one should
be careful when using CO isotopes to derive the H$_2$ mass of, e.g.,
envelopes around young stars assuming a standard abundance. With
depletion the derived H$_2$ envelope masses will be
underestimated. Another often encountered assumption, which may
introduce systematic errors, is that the lower levels are thermalised
and that the Boltzmann distribution can be used to calculate the
excitation and thus column density of a given molecular species. In
Fig.~\ref{lte}, the level populations for C$^{18}$O in the outer shell
of the model of two sources, \object{L723} and \object{TMR1}, are
shown together with the variation of the ratios of the level
populations from the Monte Carlo modelling by assuming LTE. For the
more dense envelope around the typical class 0 object, \object{L723},
the LTE assumption gives accurate results within 5--10\% in the lower
levels, but somewhat more uncertain results for the higher levels. For
the less dense envelope around \object{TMR1} (class I object) it is,
however, clear that the levels are subthermally excited and the LTE
approximation provides a poor representation of the envelope
structure.

\subsection{CO abundance jump or not?}
The large difference in CO abundance found between the class 0
sources and pre-stellar cores on the one side and the class I objects
on the other side warrants further discussion.

The apparently low CO abundances and the possible relation to
freeze-out of CO raises the question whether the assumption of a
constant fractional abundance is realistic: freezing out of pure
CO-ice and isotopes is expected to occur at roughly 20~K under
interstellar conditions \citep[e.g.,][]{sandford93}, so one would
expect to find a drastic drop in CO abundance in the outer parts of
the envelope. Due to the uncertainties in the properties of the
exterior regions, however, a change in abundance at 20~K can neither
be confirmed nor ruled out. As Table~\ref{line_int_jcmt} indicates
both the intensities of the 2--1 and 3--2 lines of C$^{18}$O and
C$^{17}$O can be fitted with a constant fractional abundance for most
sources. One can introduce strong depletion through a ``jump'' in the
fractional abundance in the region of the envelope with temperatures
lower than 20~K. This naturally leads to lower line intensities of
especially the 2--1 and 1--0 lines, but also the 3--2 lines. Since it
is mainly the 3--2 line that constrains the abundance in the inner
part, it is possible to raise the abundances of the warm regions by up
to a factor 2, if CO in the outer part is depleted by a factor of 10
or more. The modelled 1--0 and 2--1 line intensities then, however,
also become weaker, which one has to compensate for by introducing
even more cold (depleted) material outside the 10~K boundary,
accounting for 50\% or more of the observed 2--1 line emission and
almost all the 1--0 emission. 

If the dust opacity law varies with radial distance in the
envelope, increasing $\kappa_\nu$ with the higher densities, the CO
abundances in the warmer regions could be higher by a factor of
two. Combining these effects would then lower the derived CO
evaporation temperature in the envelope material towards the 20~K
evaporation temperature of pure CO ice. A changing dust opacity law
would be an interesting result in itself, which possibly can be
confirmed or disproved through modelling of other chemical species
and/or by comparing the exact line profiles with realistic models of
the velocity field. Without such an effect introduced, however, the
results presented here favors the somewhat higher evaporation
temperature for the CO.

The differences between the class 0 and I objects and a warmer
evaporation temperature may be consistent with new laboratory
experiments on the trapping and evaporation of CO on H$_2$O ice by
\cite{collings02}. Their experiments refer to amorphous H$_2$O ice
accreted layer-by-layer so that it has a porous ice structure. This
situation may be representative of the growth of H$_2$O ice layers in
pre-stellar cores and YSO envelopes. CO is deposited on top of the
H$_2$O ice. When the sample is heated from 10 to $\sim$30~K
(laboratory temperatures), some of the CO evaporates, but another
fraction diffuses into the H$_2$O ice pores. Heating to 30--70~K
allows some CO to desorb, but a restructuring of the H$_2$O ice seals
off the pores, and the remaining CO stays trapped until at least
140~K. Under interstellar conditions, the temperatures for these
processes may be somewhat lower, but it does indicate that a
significant fraction of CO can be trapped in a porous surface and that
evaporation may occur more gradually from 30~K to $\sim$60~K. A
similar property was suggested by \cite{ceccarelli01}, who found that
the dust mantles in the envelope around IRAS 16293-2422 had an
onion-like structure with H$_2$CO being trapped in CO rich ices in the
outermost regions and with the ices becoming increasingly more H$_2$O
rich when moving inwards toward higher temperatures. In this
scenario, it is not surprising that even the 3--2 lines tracing the
warmer material indicate low CO abundances. Observations of even
higher lying CO rotational lines (e.g., 6--5) would be needed probe
the full extent of this evaporation. Also infrared spectroscopy of CO
ices may reveal differences between the class 0 and I objects.

\section{Conclusion}
This is the first paper in a survey of the physical and chemical
properties of a sample of low-mass protostellar objects.  The
continuum emission from the envelopes around these objects has been
modelled using the 1D radiative transfer code, DUSTY, solving for the
temperature distribution assuming simple power-law density
distributions of the type $\rho\propto r^{-\alpha}$. For the class 0
and I objects in the sample, the brightness profiles from SCUBA 450
and 850~$\mu$m data and the SEDs from various literature studies can
be successfully modelled using this approach with $\alpha$ in the
range from 1.3 to 1.9 within $\sim$10000~AU with a typical uncertainty
of $\pm 0.2$, while it fails for the pre-stellar cores. For four
sources the profiles are indeed flatter than predicted by, e.g., the
models of \cite{shu77}, but it is argued that this could be due to
source asymmetries and/or the presence of extended cloud
material. Taking this into account, no significant difference seems to
exist between the class 0 and I sources in the sample with respect to
the shape of the density distribution, while, as expected, the class 0
objects are surrounded by more massive envelopes.

The physical models derived using this method have been applied in
Monte Carlo modelling of C$^{18}$O and C$^{17}$O data, adopting an
isothermal Bonnor-Ebert sphere as a physical model for the pre-stellar
cores. The 2--1 and 3--2 lines can be modelled for all sources with
constant fractional abundances of the isotopes with respect to H$_2$
and an isotope ratio [$^{18}$O/$^{17}$O] of 3.9, in agreement with the
``standard'' value for the local interstellar medium. The 1--0 lines
intensities, however, are significantly underestimated in the models
compared to the observations, indicating that ambient cloud emission
contributes significantly or that the outer parts of the envelopes are
not well accounted for by the models. The derived abundances increase
with decreasing envelope mass - with an average CO abundance of
$2.0\times 10^{-5}$ for the class 0 objects and pre-stellar cores, and
$1.1\times 10^{-4}$ for the class I objects. The 3--2 lines indicate
that the lower CO abundance in class 0 objects also applies to the
regions of the envelopes with temperatures higher than $\sim 20-25$~K,
the freeze-out of pure CO ice. This feature can be explained if a
significant fraction of the solid CO is bound in a (porous) ice
mixture from where it does not readily evaporate. The physical models
presented here will form the basis for further chemical modelling of
these sources.

\begin{acknowledgements}
The authors thank Michiel Hogerheijde and Floris van der Tak for use
of their Monte Carlo code and stimulating discussions, and Helen
Fraser for sharing the results of the new experiments on CO
evaporation prior to publication. They are grateful to Sebastien Maret
for carrying out the IRAM observations in November 2001. The referee,
Cecilia Ceccarelli, is thanked for thorough and insightful comments,
which helped in clarifying the results of the paper. The work of JKJ
is funded by the Netherlands Research School for Astronomy (NOVA)
through a \emph{netwerk 2, Ph.D. stipend}. FLS is funded by grants
from the University of Leiden and the Netherlands Foundation for
Scientific Research (NWO) no.\ 614.041.004. Astrochemistry in Leiden
is supported by a Spinoza grant. This article made use of data
obtained through the JCMT archive as Guest User at the Canadian
Astronomy Data Center, which is operated by the Dominion Astrophysical
Observatory for the National Research Council of Canada's Herzberg
Institute of Astrophysics. The SIMBAD database, operated at CDS,
Strasbourg, France has been intensively used as well.
\end{acknowledgements}

\end{document}